\documentclass[12pt]{article}

\usepackage{tikz}
\usetikzlibrary{calc}
\usepackage{pgfplots}
\pgfplotsset{compat=1.18}

\usepackage[margin=1in]{geometry}
\usepackage{amsmath, amssymb, amsthm}

\usepackage{graphicx}
\usepackage{hyperref}
\usepackage{mathrsfs}
\usepackage{enumitem}
\usepackage{longtable}
\usepackage{booktabs}

\usepackage{fancyhdr}
\usepackage{setspace}
\usepackage{titlesec}
\usepackage{authblk}

\titleformat{\section}{\normalfont\Large\bfseries}{\thesection}{1em}{}
\titleformat{\subsection}{\normalfont\large\bfseries}{\thesubsection}{1em}{}

\title{\textbf{The Autonomy of the Lightning Network: \\A Mathematical and Economic Proof of Structural Decoupling from BTC}}
\author{Craig Wright}
\date{}

\begin{document}
\maketitle

\begin{abstract}
\noindent This paper presents a rigorous formal and game-theoretic analysis of layered transaction networks, focusing on the Lightning Network (LN) as an off-chain overlay to Bitcoin’s constrained base layer (BTC). We develop a comprehensive mathematical framework incorporating axiomatic definitions, asymptotic cost models, and agent-based strategic dynamics. Beginning from foundational constraints in on-chain throughput and off-chain liquidity routing, we construct formal proofs outlining the economic infeasibility of long-term settlement on BTC and the emergence of rent-extracting hub hierarchies within LN.

\noindent We show that as transaction demand grows, BTC’s cost structure becomes asymptotically prohibitive due to bounded blockspace, while LN’s costs flatten, favouring oligopolistic routing infrastructure. Using game-theoretic constructs and strategic equilibrium analysis, we prove that Lightning hubs evolve into quasi-monopolistic entities exerting control over liquidity and routing paths, creating closed-loop rent cycles and systemic centralisation. We analyse Lightning’s evolution into a shadow banking system, demonstrating parallels to historical financial collapse scenarios rooted in opacity, leverage, and regulatory absence.

\noindent Furthermore, we extend the analysis into computational complexity theory. Drawing on Even et al.\ (1975) and Wright (2023), we formalise the NP-completeness of decentralised route selection under liquidity constraints, and show that as system size increases, real-time routing becomes computationally intractable without centralised heuristics. This leads to an unavoidable tension: decentralised topology cannot scale under strict complexity bounds, resulting in economic and infrastructural centralisation as an emergent inevitability.

\noindent Finally, we develop proofs, lemmas, and diagrammatic models of cost asymptotics, liquidity control, strategic dominance, and structural disconnection between base-layer finality and overlay transaction execution. Our results demonstrate that LN does not solve the scalability problem but instead transforms it into a form of economic enclosure, displacing decentralisation with structural rent-seeking and liquidity capture.

\end{abstract}

\begin{center}
\end{center}

\newpage

\tableofcontents
\newpage

\section{Introduction}

The Bitcoin protocol was initially designed as a peer-to-peer digital cash system with a simple economic premise: transfer of value directly between users without reliance on centralised intermediaries. However, inherent constraints in on-chain throughput, particularly the limited block size and fixed block interval, impose strict scalability bottlenecks in the BTC fork of Bitcoin. To overcome these, the Lightning Network (LN) was introduced as a layered solution, routing payments off-chain via bidirectional payment channels. While initially presented as a decentralised scaling enhancement, LN alters the economic topology of the system, transforming the infrastructure into a strategic rent-extraction network.

This paper presents a unified analytical framework that formalises the structural, economic, and computational consequences of layered transaction architectures. Our approach synthesises mathematical modelling, game-theoretic reasoning, and computational complexity analysis to demonstrate the systemic evolution of LN into a closed, quasi-monopolistic overlay. We begin by defining a dual-layered system structure, formally articulating the economic roles of agents, routing functions, and liquidity distributions. Axiomatic constraints—such as throughput ceilings and positive transactional utility—are used to derive asymptotic cost models for both BTC and LN, highlighting divergent behaviours under demand escalation.

We introduce a formal set of proofs showing that the Lightning Network does not merely inherit but amplifies centralising tendencies through strategic liquidity consolidation. Hubs emerge as dominant agents in an effectively rentier economy, reintroducing structural chokepoints that the original protocol sought to eliminate. Diagrams illustrate these dynamics: from topological control loops to asymptotic cost differentials and disconnected settlement layers.

The model is further extended through the lens of computational complexity. Building on Even, Itai, and Shamir (1975), we show that the problem of routing payments under liquidity constraints is NP-complete. Wright (2023) provides a parallel argument, demonstrating that layered payment systems converge to deterministic strategic equilibria only under centralised assumptions, undermining real-time decentralised scalability.

The goal is to make these conclusions explicit through rigorous, formal derivation. We identify Lightning not as a neutral scaling layer, but as a closed, rent-extracting financial overlay that mimics and inherits the systemic risks of traditional shadow banking. We conclude with policy implications, structural critiques, and design alternatives rooted in scalable, transparent economic architectures.

\subsection{Motivation and Scope}

The prevailing narrative surrounding the Lightning Network promotes it as a scalable, decentralised solution to Bitcoin’s base-layer limitations. However, this interpretation often omits the underlying economic and computational transformations that arise from layering. As the Bitcoin base chain remains constrained by a hard cap on block size and fixed intervals, demand pressure shifts economic weight toward routing intermediaries in the overlay system. In this context, liquidity is no longer merely a passive medium—it becomes a competitive resource whose distribution determines systemic influence and strategic power.

This paper is motivated by the need to rigorously articulate the implications of this shift. We examine how cost escalations on the base layer incentivise migration to a layered topology that is fundamentally non-neutral. We model how transaction cost asymptotics, strategic agent behaviour, and liquidity centralisation interact to create emergent monopolistic structures. Through formal mathematical exposition and computational proofs, we explore the intrinsic limitations of decentralised routing and liquidity reallocation under bounded rationality and economic pressure.

The scope of this work encompasses three domains: (1) formal economic modelling of transaction networks under constrained and layered architectures; (2) game-theoretic evaluation of agent strategies within Lightning-style liquidity networks; and (3) computational analysis of payment routing as a complexity-theoretic optimisation problem. The Lightning Network is presented not as an isolated construct but as a paradigm of layered infrastructure subject to economic rent extraction and structural divergence from Bitcoin’s foundational principles.

By unifying these perspectives, we offer a new analytical toolkit for evaluating not only LN but the broader class of financial overlays masquerading as neutral scalability solutions. Our results apply generally to any architecture that introduces closed, settlement-delayed layers above a constrained consensus mechanism, and thus bear relevance for system architects, economic theorists, and regulatory stakeholders alike.

\subsection{BTC, the Lightning Network, and the Settlement Problem}

Bitcoin’s original design as a digital cash system was centred on direct, on-chain settlement. However, BTC, the protocol maintained by the BTC Core developers, enforces severe throughput constraints—primarily via the 1MB block size limit and an enforced ten-minute block interval. These protocol-level restrictions create an artificial scarcity of block space, thereby elevating transaction fees under demand pressure and constraining the scalability of direct settlement.

The Lightning Network (LN) has been proposed as a solution to this bottleneck by shifting transactional load off-chain. In principle, LN allows parties to open payment channels via on-chain commitments, conduct numerous off-chain transactions, and settle only the net result back onto the chain. While this design promises high throughput and low latency, it introduces a fundamental shift in settlement semantics: instead of immediate global finality, LN relies on contingent, revocable commitments enforced by time locks and penalty mechanisms.

This settlement model introduces several critical challenges. Firstly, finality in Lightning is conditional, not absolute. The security of funds depends on the constant availability and monitoring of channel states, either by the user or a delegated watchtower. Secondly, off-chain transaction records are not globally visible, making auditing and forensic analysis practically impossible. Thirdly, channel closures—especially in high-volume or contested environments—are increasingly expensive and time-constrained, undermining the utility of the base chain as a universal recourse mechanism.

As adoption grows, LN’s reliance on limited on-chain settlement capacity becomes a structural bottleneck. The very mechanism meant to alleviate congestion becomes dependent on the scarce resource it seeks to bypass. Settlement becomes probabilistic and exclusionary, driven by topology, liquidity availability, and temporal constraints. Consequently, the Lightning Network does not provide scalable final settlement in the same sense as Bitcoin’s base layer; rather, it constitutes a credit-based system with delayed and contingent resolution, structurally akin to historical clearinghouses or modern shadowbanking infrastructure.

In this section, we formalise this divergence and establish the systemic consequences of displacing final settlement into contingent off-chain layers.

\subsection{Outline of the Argument and Method}

This paper presents a formal, axiomatic, and game-theoretic critique of the BTC protocol’s reliance on the Lightning Network (LN) as a scalability mechanism. Our argument proceeds in several stages, each integrating tools from theoretical computer science, network economics, automata theory, and macro-financial systems modelling. The central thesis is that the introduction of off-chain routing structures fundamentally alters both the semantic and economic structure of Bitcoin, transforming it from a system of direct digital cash settlement into a probabilistic, credit-based, rent-extracting overlay—resembling centralised financial intermediaries more than a disintermediated ledger.

We begin by establishing a rigorous formal system \(\mathscr{S}\), composed of the base layer ledger \(\mathscr{B}\), the overlay payment network \(\mathscr{L}\), the set of rational agents \(\mathscr{A}\), and relevant utility, cost, and time-indexed state transitions. Using this system, we construct a series of axioms capturing the structural limitations of BTC’s base layer, the unbounded transactional demand in real-world systems, and the emergent optimisation behaviour of rational agents under cost constraints.

Subsequent sections model the escalating cost structure of BTC under congestion, using empirical and asymptotic analysis to establish an exponential fee-growth trajectory. We contrast this with Lightning’s near-flat cost asymptote, under ideal liquidity assumptions, and demonstrate via TikZ-illustrated diagrams and functional proofs the resulting divergence in network behaviour. Strategic agent models are introduced to analyse emergent oligopoly in liquidity hubs, capturing their role as shadowbanks and rent-extracting intermediaries under liquidity centralisation pressures.

From there, we proceed to construct a game-theoretic topology, showing how routing control becomes a dominant strategy and how equilibrium fee structures under different topologies lead to rent-seeking and systemic exclusion. The discussion extends to macroeconomic parallels, showing how Lightning replicates key features of shadow financial systems, including opacity, synthetic monetary bases, and absence of reserve discipline.

We then derive equilibrium conditions, present asymptotic bounds, and provide structural proofs that Lightning cannot sustain scalable settlement without violating core design principles of Bitcoin. We conclude by articulating the implications for network design, demonstrating that base-layer scaling is not optional but structurally essential if settlement finality is to be retained.

This approach allows us to integrate computational complexity bounds (Even \& Itai, 1975), economic topology (Barabási, 1999), and formal systems modelling (Wright, 2023) into a unified critique that is both mathematically rigorous and economically grounded.

\section{Axiomatic Foundations}

The analytical rigour of this paper rests upon a clearly articulated axiomatic framework. In constructing a formal economic and game-theoretic proof of the Lightning Network's structural separation from BTC, we require an explicit delineation of the system's operational boundaries and logical commitments. This section serves as the bedrock for all subsequent argumentation, modelling the behaviour of rational agents interacting under constraints imposed by protocol design, market incentives, and capacity limitations.

We begin by specifying the structural system we are analysing. This includes the distinction between on-chain settlement via the BTC blockchain and off-chain transactional clearing within the Lightning Network. The system must be framed such that each element—transactions, fees, liquidity, routing agents, and capacity ceilings—can be captured through formal symbols and subjected to mathematical reasoning.

We then assert a minimal but complete set of axioms. These axioms describe systemic invariants: upper bounds on BTC transaction throughput, the potential unboundedness of off-chain demand, the non-zero utility of economic exchanges, and the emergent liquidity consolidation that arises from cost-minimising rational behaviour. These axioms are neither conjectural nor empirical in nature; they are formal constraints embedded in the logic of the design or extrapolated from economic first principles.

The final subsection of this section introduces the formal notational system. This includes variable conventions and mathematical functions that govern all cost functions, equilibrium outcomes, and limit expressions used throughout the paper. The notation is chosen to prioritise clarity, parsimony, and interoperability across disciplines, ensuring that mathematical, economic, and computational interpretations of the model can be unified within a single formal grammar.

The subsections that follow detail the definitional scope, enumerate the axioms in precise form, and lay out the symbolic architecture that enables a rigorous derivation of our core theorems.

\subsection{System Definitions}

To establish a rigorous analytical framework, we define the computational-economic system under study as a dual-layered transactional network composed of a constrained base ledger and a scalable overlay. The formal system is described by the septuple:

\[
\mathscr{S} = (\mathscr{B}, \mathscr{L}, \mathscr{A}, \mathscr{R}, \mathscr{V}, \mathscr{C}, \mathscr{T})
\]

\noindent where:

\begin{enumerate}[label=(\roman*)]
    \item \( \mathscr{B} \) is the base-layer system, instantiated as a blockchain ledger with consensus finality, throughput constraint, and global visibility.
    \item \( \mathscr{L} \) is the overlay system, here instantiated as the Lightning Network, modelled as a dynamic, weighted, directed graph of bi-directional payment channels between rational agents.
    \item \( \mathscr{A} \) is the set of agents, partitioned into users \( \mathcal{U} \), liquidity hubs \( \mathcal{H} \subset \mathcal{U} \), watchtowers \( \mathcal{W} \), and miners \( \mathcal{M} \), each with distinct operational strategies and state transition functions.
    \item \( \mathscr{R} \) is the set of routing functions mapping transaction intents to feasible payment paths in \( \mathscr{L} \) under liquidity and timelock constraints.
    \item \( \mathscr{V} \subset \mathbb{R}_{>0} \) is the set of transaction utility values. Every transaction has a non-zero valuation, interpreted as subjective economic utility and denominated in satoshis.
    \item \( \mathscr{C} \) is the system-wide cost function space, combining both base-layer and overlay costs as measurable penalty functions over temporal epochs.
    \item \( \mathscr{T} \subset \mathbb{N} \) is the discrete temporal index space, representing time as a sequence of atomic epochs indexed \( t \in \mathscr{T} \).
\end{enumerate}

\paragraph{Base Layer \(\mathscr{B}\):} Let \( \mathscr{B} = \langle \mathcal{B}, \sqsubseteq, \delta \rangle \) where:
\begin{itemize}[label=--]
    \item \( \mathcal{B} \) is the set of blocks, each block \( b_i \in \mathcal{B} \) encoding a finite sequence of transactions.
    \item \( \sqsubseteq \) is the total order imposed by Nakamoto consensus under Proof-of-Work, forming a chain \( \mathcal{B}_\text{chain} \subseteq \mathcal{B} \) such that \( \forall i < j,\; b_i \sqsubset b_j \).
    \item \( \delta: \mathscr{T} \to \mathbb{N} \) is the block interval function, determining the temporal separation between valid blocks, constrained by protocol to approximate a fixed \( \Delta t \in \mathbb{R}_{>0} \).
\end{itemize}

The block size limit \( s \) and interval \( \Delta t \) induce an upper bound on transaction throughput \( T_{\max} \), such that:

\[
T_{\max} = \left\lfloor \frac{s}{\bar{t}_{\text{tx}} \cdot \Delta t} \right\rfloor
\]

\noindent where \( \bar{t}_{\text{tx}} \) is the average transaction size. Empirical measurements suggest this limit rarely exceeds 5 transactions per second, even under Segregated Witness enhancements and transaction batching \cite{croman2016scaling,poon2016bitcoin}.

\paragraph{Overlay Network \(\mathscr{L}\):} Let \( \mathscr{L}(t) = \langle V(t), E(t), \lambda(t) \rangle \) be the overlay network at time \( t \), where:
\begin{itemize}[label=--]
    \item \( V(t) \subseteq \mathscr{A} \) is the set of active nodes (agents) participating in Lightning at time \( t \).
    \item \( E(t) \subseteq V(t) \times V(t) \) is the set of directed, bi-directional payment channels between nodes.
    \item \( \lambda: E(t) \to \mathbb{R}_{\geq 0} \times \mathbb{R}_{\geq 0} \) is the liquidity map such that for channel \( e_{ij} \), \( \lambda(e_{ij}) = (\ell_{ij}, \ell_{ji}) \) denotes available outbound balances in each direction.
\end{itemize}

Channel creation and closure require base-layer transactions. In practice, however, Lightning assumes that the vast majority of transactions are routed off-chain, never necessitating settlement in \( \mathscr{B} \) \cite{roos2019settling}.

\paragraph{Agents \(\mathscr{A}\):} Each agent \( a_k \in \mathscr{A} \) is a computational entity defined by a strategy function \( \sigma_k: \mathscr{S} \to \mathbb{R} \) that maximises discounted utility:

\[
\max_{\sigma_k} \mathbb{E} \left[ \sum_{t = 0}^{\infty} \delta^t u_k(t) \right]
\]

\noindent where \( u_k(t) \) is the agent’s utility at time \( t \), and \( \delta \in (0, 1] \) is the discount factor. Strategies span route selection, fee management, liquidity optimisation, and adversarial deviation, subject to system rules.

\paragraph{Routing \(\mathscr{R}\):} The routing layer is defined as a function \( \rho: \mathcal{P} \to \{0,1\} \), where \( \mathcal{P} \) is the set of all simple paths between source and destination nodes in \( \mathscr{L} \). The function \( \rho \) returns 1 if the path satisfies atomicity, timelock, and liquidity constraints, and 0 otherwise. Let each feasible path be scored by a path-cost functional \( \phi(P) \), with optimal paths selected by:

\[
P^* = \arg \min_{P \in \mathcal{P},\; \rho(P) = 1} \phi(P)
\]

Pathfinding in \( \mathscr{L} \) is computationally non-trivial due to the non-global knowledge of liquidity states and the requirement for atomic multi-hop execution \cite{malavolta2017concurrency}.

\paragraph{Transaction Utility \(\mathscr{V}\):} Each transaction \( \tau \) is associated with a valuation \( v(\tau) \in \mathscr{V} \), where \( \mathscr{V} \subset \mathbb{R}_{>0} \). We assume quasi-linearity: utility from a payment is linear in \( v \), and independent of concurrent transactions. For the transaction set \( \Pi(t) = \{\tau_i\} \) active at time \( t \), the aggregate system utility is:

\[
U_{\text{net}}(t) = \sum_{\tau \in \Pi(t)} v(\tau)
\]

\paragraph{Cost Functions \(\mathscr{C}\):} Let \( \mathscr{C} = \{ c_i: \mathscr{S} \to \mathbb{R}_{\geq 0} \} \), where each \( c_i \) maps a transaction configuration to a cost. Define two distinct subspaces:

\begin{align*}
\mathscr{C}_{\mathscr{B}} &= \text{costs from on-chain congestion, delay, and fee escalation}, \\
\mathscr{C}_{\mathscr{L}} &= \text{costs from routing fees, liquidity lockup, and failure probabilities}.
\end{align*}

\paragraph{Temporal Indexing \(\mathscr{T}\):} Time is modelled as a discrete sequence \( t \in \mathscr{T} \), driving all dynamic system variables. The base layer evolves monotonically via block additions, while the overlay evolves stochastically as agents act.

This formal structure underpins the axiomatic framework of the paper. Each axiom in the next subsection draws on one or more components of \( \mathscr{S} \), ensuring precision and coherence in our theoretical derivations.

\subsection{Axioms}

This subsection establishes the foundational axioms that govern the behaviour of the dual-layered transactional system \( \mathscr{S} \). These axioms are defined categorically and serve as immutable logical commitments for all subsequent propositions, lemmas, and theorems. Each axiom is framed to reflect inherent structural, economic, and algorithmic characteristics of the base and overlay layers. They are internally consistent, irreducible, and together jointly sufficient to derive the equilibrium outcome of Lightning Network autonomy from the BTC settlement layer.

Let all axioms be globally quantified over time \( t \in \mathscr{T} \), and let all functions and structures refer to their definitions in the system tuple \( \mathscr{S} = (\mathscr{B}, \mathscr{L}, \mathscr{A}, \mathscr{R}, \mathscr{V}, \mathscr{C}, \mathscr{T}) \).

\paragraph{Axiom 1 (On-Chain Throughput Constraint).}  
The BTC base layer \( \mathscr{B} \) is constrained by a global, time-invariant upper bound on transaction throughput:

\[
\exists T_{\max} \in \mathbb{N},\; \forall t \in \mathscr{T},\; |\Pi_{\mathscr{B}}(t)| \leq T_{\max}
\]

\noindent where \( \Pi_{\mathscr{B}}(t) \subseteq \Pi(t) \) is the set of transactions submitted to \( \mathscr{B} \) at time \( t \). This constraint is imposed by the protocol parameters of BTC: block size \( s \), average transaction size \( \bar{t}_{\text{tx}} \), and inter-block interval \( \Delta t \). As of the reference configuration (SegWit-enabled Bitcoin Core), the empirical upper bound is approximately \( T_{\max} \approx 5 \) transactions per second. This condition is invariant under block congestion and independent of transaction demand.

\paragraph{Axiom 2 (Unbounded Off-Chain Transaction Demand).}  
There is no intrinsic upper bound on the demand for off-chain transactions routed through the Lightning Network \( \mathscr{L} \):

\[
\forall n \in \mathbb{N},\; \exists t \in \mathscr{T} \text{ such that } |\Pi_{\mathscr{L}}(t)| > n
\]

\noindent where \( \Pi_{\mathscr{L}}(t) \subseteq \Pi(t) \) denotes the set of transactions routed exclusively through the overlay network \( \mathscr{L} \). This axiom reflects an assumption of economic scalability: as long as agents remain incentivised to transact and the value of transactions remains positive, demand for off-chain transactions may grow without bound, subject only to liquidity availability and path feasibility. This growth trajectory is assumed to follow an open economic system with unbounded potential transactional intensity.

\paragraph{Axiom 3 (Transactional Utility Positivity).}  
Each transaction executed within \( \mathscr{S} \) confers a strictly positive utility to the transacting agents:

\[
\forall \tau \in \Pi(t),\; v(\tau) \in \mathscr{V} \Rightarrow v(\tau) > 0
\]

\noindent where \( v: \Pi \to \mathscr{V} \subset \mathbb{R}_{>0} \) is the transaction valuation function. This axiom encodes a basic rationality principle from microeconomics: voluntary transactions are executed only when they yield nonzero economic benefit to both parties. In the Lightning Network specifically, this assumption guarantees that fee-bearing paths can exist without arbitrage against the base layer, provided costs remain below valuation.

\paragraph{Axiom 4 (Liquidity Consolidation under Cost Minimisation).}  
Under repeated execution of routing and rebalancing algorithms in \( \mathscr{L} \), the system tends toward topologies in which liquidity is centralised in a minimal subset of nodes:

\[
\exists \mathcal{H}_\infty \subset \mathcal{H} \text{ with } |\mathcal{H}_\infty| \ll |\mathcal{H}| \text{ such that } \lim_{t \to \infty} \frac{\sum_{h \in \mathcal{H}_\infty} \lambda_h(t)}{\sum_{h \in \mathcal{H}} \lambda_h(t)} = 1
\]

\noindent where \( \lambda_h(t) \) is the total active liquidity held by hub \( h \) at time \( t \). This is a formalisation of the emergence of "superhubs" or shadowbank entities within Lightning, driven by the minimisation of routing cost functionals and reinforcement of high-throughput paths. This axiom is supported by observable network evolution in similar scale-free and preferential attachment structures, including empirical studies of Lightning's topology under increasing usage.

These four axioms jointly ground the system in computational and economic constraints, provide an upper limit on base-layer participation, and establish the conditions under which routing control converges into dominant central nodes. From these premises, we derive the necessary conditions for the economic and structural decoupling of the Lightning Network from the BTC base chain in the subsequent sections.

\subsection{Formal Notation and Variable Definitions}

This subsection formally introduces the symbolic and notational conventions employed throughout the remainder of the paper. All variables, functionals, operators, and set-theoretic constructs are defined in strict correspondence with the system model \( \mathscr{S} = (\mathscr{B}, \mathscr{L}, \mathscr{A}, \mathscr{R}, \mathscr{V}, \mathscr{C}, \mathscr{T}) \), previously established in the system definitions. The notation is designed to ensure algebraic tractability, semantic clarity, and logical determinism across mathematical derivations, economic equilibria, and asymptotic constructions.

\paragraph{Sets and Structures}

\begin{itemize}[label=--]
  \item \( \mathscr{B} \): The base-layer blockchain ledger, with block set \( \mathcal{B} = \{ b_i \} \), ordering \( \sqsubseteq \), and time-separation map \( \delta: \mathscr{T} \to \mathbb{N} \).
  \item \( \mathscr{L} \): The Lightning Network overlay, modelled at time \( t \) as \( \mathscr{L}(t) = (V(t), E(t), \lambda(t)) \).
  \item \( \mathscr{A} \): The agent space, partitioned as:
    \[
    \mathscr{A} = \mathcal{U} \cup \mathcal{H} \cup \mathcal{M} \cup \mathcal{W}
    \]
    where \( \mathcal{U} \) = end-users, \( \mathcal{H} \) = liquidity hubs, \( \mathcal{M} \) = miners, \( \mathcal{W} \) = watchtowers.
  \item \( \mathscr{R} \): The routing function set, where each \( \rho \in \mathscr{R} \) is a map:
    \[
    \rho: \mathcal{P} \to \{0,1\}
    \]
    indicating feasibility of a path \( P \) within \( \mathscr{L}(t) \).
  \item \( \mathscr{V} \subset \mathbb{R}_{>0} \): The value space of transactions.
  \item \( \mathscr{C} \): The cost function space, partitioned:
    \[
    \mathscr{C} = \mathscr{C}_{\mathscr{B}} \cup \mathscr{C}_{\mathscr{L}}
    \]
  \item \( \mathscr{T} \subset \mathbb{N} \): Discrete time index set.
\end{itemize}

\paragraph{Functions and Variables}

\begin{itemize}[label=--]
  \item \( t \in \mathscr{T} \): Global time index.
  \item \( \Pi(t) \): Set of all transactions occurring at time \( t \).
  \item \( \Pi_{\mathscr{B}}(t), \Pi_{\mathscr{L}}(t) \): Transactions processed via base layer and Lightning, respectively.
  \item \( \lambda(e_{ij}) = (\ell_{ij}, \ell_{ji}) \): Liquidity state of a bidirectional channel \( e_{ij} \in E(t) \).
  \item \( \tau \in \Pi(t) \): A single transaction.
  \item \( v(\tau) \in \mathscr{V} \): Utility of transaction \( \tau \), with \( v: \Pi(t) \to \mathscr{V} \).
  \item \( u_k(t) \): Instantaneous utility for agent \( a_k \) at time \( t \).
  \item \( \sigma_k: \mathscr{S} \to \mathbb{R} \): Strategy function for agent \( a_k \in \mathscr{A} \).
  \item \( T_{\max} \in \mathbb{N} \): Maximum number of on-chain transactions per second.
  \item \( s \in \mathbb{N} \): Maximum block size in bytes.
  \item \( \bar{t}_{\text{tx}} \in \mathbb{R}_{>0} \): Average transaction size.
  \item \( \Delta t \in \mathbb{R}_{>0} \): Inter-block time interval.
  \item \( \phi(P) \): Path cost functional over Lightning paths.
  \item \( F_h(t) \in \mathbb{R}_{\geq 0} \): Routing fee charged by hub \( h \in \mathcal{H} \) at time \( t \).
  \item \( R_h(t) \): Revenue of hub \( h \), typically:
    \[
    R_h(t) = F_h(t) \cdot |\Pi_h(t)|
    \]
    where \( \Pi_h(t) \subseteq \Pi_{\mathscr{L}}(t) \) are the transactions routed through hub \( h \).
\end{itemize}

\paragraph{Operators and Symbols}

\begin{itemize}[label=--]
  \item \( \sqsubseteq \): Block ordering under longest valid chain consensus.
  \item \( \cdot \): Scalar multiplication in cost and revenue terms.
  \item \( \arg\min, \arg\max \): Optimal selection operators over feasible path sets.
  \item \( \lim_{t \to \infty} \): Asymptotic temporal limit, used in liquidity convergence and divergence of fees.
  \item \( \forall, \exists \): Universal and existential quantifiers as used in formal axioms.
\end{itemize}

The above definitions provide the complete notational basis for the remainder of the paper. All forthcoming equations, proofs, and functional dependencies are derived using only these objects and their permitted compositions. The notation is closed, unambiguous, and sufficient to specify discrete-time economic games over the dual-layer payment infrastructure \( \mathscr{S} \).

\section{Network Cost Structures}

This section introduces the formal construction of network-wide cost structures that govern the decision-making logic of agents within the system \( \mathscr{S} \). Specifically, we develop cost functionals associated with two transactional pathways: (i) the base-layer blockchain \( \mathscr{B} \), and (ii) the overlay Lightning Network \( \mathscr{L} \). These structures are not merely accounting devices but form the economic substrate of the network’s evolution, as they induce incentive gradients, equilibrium states, and strategic behaviour among rational agents.

We begin by formalising the transaction cost incurred on the base chain. The cost is primarily a function of mempool congestion, constrained block capacity, and auction-based miner selection. Under a bounded throughput regime, this cost exhibits nonlinear behaviour in the presence of excess demand. We define this structure in terms of a demand-dependent penalty function \( C_{\mathscr{B}}(t) \), which quantifies the marginal cost per transaction as a function of total system demand \( |\Pi(t)| \) relative to the throughput ceiling \( T_{\max} \).

In contrast, the Lightning Network \( \mathscr{L} \) exhibits a different cost topology. Transaction costs here are governed not by global consensus or miner selection, but by the liquidity topology of \( \mathscr{L}(t) \), the local routing preferences of hubs, and the relative pricing power of nodes in control of highly connected channels. These costs are captured by a routing cost functional \( C_{\mathscr{L}}(t) \), defined over feasible paths \( P \in \mathcal{P} \) and modulated by routing fees \( F_h(t) \), liquidity penalties, and timelock exposure. Unlike the base-layer cost, these costs are continuous, path-dependent, and often sublinear in high-liquidity regimes.

This section is structured as follows. We first define and characterise the BTC base-layer transaction cost function, including the derivation of congestion multipliers and asymptotic divergence. We then develop the Lightning routing cost framework, including the formalisation of path-based fee accumulation and liquidity-dependent routing friction. Next, we provide asymptotic comparisons between base-layer and overlay-layer costs under increasing transactional demand. Finally, we establish a threshold condition under which Lightning transactions become strictly dominant in cost efficiency over any finite time horizon, assuming liquidity continuity.

Each component of the cost structure is embedded within the symbolic formalism established earlier, and all results are presented with mathematical rigour suitable for equilibrium derivation in later game-theoretic modelling. The section also includes graphical representations of cost divergence, which will be extended in structural separation theorems and policy risk modelling in subsequent sections.
\subsection{BTC Transaction Fee Escalation}

We formalise the cost of transacting on the BTC base layer \( \mathscr{B} \) as a function of constrained capacity and auction-based fee dynamics. The BTC network imposes a hard upper limit on the number of transactions that can be included in each block, defined by the block size \( s \), average transaction size \( \bar{t}_{\text{tx}} \), and the inter-block interval \( \Delta t \). From this, the maximum effective transaction throughput \( T_{\max} \in \mathbb{N} \) is given by:

\[
T_{\max} = \left\lfloor \frac{s}{\bar{t}_{\text{tx}} \cdot \Delta t} \right\rfloor
\]

Let \( D(t) = |\Pi(t)| \) be the total transaction demand at time \( t \). The key structural constraint is:

\[
\forall t \in \mathscr{T},\quad |\Pi_{\mathscr{B}}(t)| \leq T_{\max}
\]

The BTC fee market operates as a priority-based auction: transactions offering higher fees per byte are selected for inclusion by miners. In equilibrium, the marginal transaction that enters the block pays a fee at or near the threshold that displaces the lowest-ranked transaction in the mempool.

We define the base-layer transaction cost function \( C_{\mathscr{B}}: \mathbb{N} \times \mathbb{N} \to \mathbb{R}_{\geq 0} \) as:

\[
C_{\mathscr{B}}(D, T_{\max}) = f_0 \cdot \max\left(1,\ \frac{D}{T_{\max}}\right)^\gamma
\]

where:
\begin{itemize}[label=--]
  \item \( f_0 \in \mathbb{R}_{> 0} \) is the baseline (non-congested) transaction fee.
  \item \( \gamma > 1 \) captures the elasticity of fee escalation under congestion.
\end{itemize}

This function reflects the convexity of the BTC fee structure under excess demand. When \( D \leq T_{\max} \), fees are stable at \( f_0 \); when \( D > T_{\max} \), fees escalate superlinearly.

\paragraph{Asymptotic Behaviour}

Letting demand grow without bound:

\[
\lim_{D \to \infty} C_{\mathscr{B}}(D, T_{\max}) = \infty
\]

This divergence establishes the following key result:

\textbf{Proposition.} In the absence of corresponding increases in throughput \( T_{\max} \), the cost of transacting on-chain under sustained excess demand becomes unbounded.

\paragraph{Implications}

This dynamic ensures that low-value or microtransactions are priced out of the BTC base layer during congestion. It also means that fee rationality enforces a de facto exclusion mechanism, in which only transactions with economic value exceeding the marginal cost will execute on-chain. Consequently, this fee structure directly incentivises agents to seek alternative execution pathways—namely, the Lightning Network \( \mathscr{L} \)—as transaction demand exceeds settlement capacity.

Subsequent subsections will formalise this redirection through a comparative analysis of routing cost in \( \mathscr{L} \), leading to the formulation of crossover conditions where off-chain pathways strictly dominate on-chain settlement in expected cost terms.

\subsection{Lightning Routing Cost Model}

In contrast to the escalating fee dynamics of the base layer \( \mathscr{B} \), the Lightning Network \( \mathscr{L} \) enables transaction execution through off-chain payment channels, where routing costs are governed by both economic and topological constraints. This subsection formalises the microeconomic cost structure of routing in \( \mathscr{L} \), grounded in the dynamic liquidity state and graph-theoretic path selection mechanisms.

Let \( G(t) = \langle V(t), E(t) \rangle \) represent the directed multigraph of active Lightning nodes and channels at time \( t \). Each edge \( e_{ij} \in E(t) \) has associated liquidity \( \lambda(e_{ij}) = (\ell_{ij}, \ell_{ji}) \) and a routing fee function:

\[
f_{ij}(x) = \alpha_{ij} + \beta_{ij} \cdot x
\]

\noindent where:
\begin{itemize}[label=--]
  \item \( x \in \mathbb{R}_{> 0} \) is the payment amount,
  \item \( \alpha_{ij} \in \mathbb{R}_{\geq 0} \) is the fixed base fee for using the channel \( e_{ij} \),
  \item \( \beta_{ij} \in \mathbb{R}_{\geq 0} \) is the proportional fee rate per satoshi transferred.
\end{itemize}

Each routed transaction \( \tau \) corresponds to a path \( P = (v_1, \dots, v_k) \subset G(t) \) from sender \( v_1 \) to receiver \( v_k \), where \( \forall i < k,\; (v_i, v_{i+1}) \in E(t) \) and \( \lambda(v_i, v_{i+1}) \geq x + \varepsilon \) for some settlement margin \( \varepsilon > 0 \). The total path cost is given by:

\[
C_{\mathscr{L}}(P, x) = \sum_{i = 1}^{k - 1} \left( \alpha_{i, i+1} + \beta_{i, i+1} \cdot x \right)
\]

Additionally, the path selection process introduces computational cost \( \mathcal{C}_{\text{search}}(P) \), which may be bounded or unbounded depending on graph sparsity, privacy constraints (e.g. onion routing), and adversarial uncertainty. Hence, the expected transaction cost is:

\[
\mathbb{E}\left[C_{\mathscr{L}}(\tau)\right] = C_{\mathscr{L}}(P^*, x) + \mathcal{C}_{\text{search}}(P^*)
\]

\noindent where \( P^* \) is the optimal feasible path with respect to some utility minimisation criterion under liquidity, cost, and failure probability.

\paragraph{Liquidity Fragmentation Cost}

One of the key structural inefficiencies in Lightning arises from liquidity fragmentation. While the network may have sufficient total liquidity \( L = \sum_{e \in E(t)} \lambda(e) \), routing large payments remains infeasible if no individual path can support the full payment value. Define the fragmentation penalty \( \psi(x) \) as:

\[
\psi(x) = \inf_{P \in \mathcal{P}_x} C_{\mathscr{L}}(P, x) - \inf_{P \in \mathcal{P}_{\infty}} C_{\mathscr{L}}(P, x)
\]

\noindent where \( \mathcal{P}_x \) is the set of feasible paths for payment size \( x \), and \( \mathcal{P}_{\infty} \) assumes infinite liquidity. As shown by Malavolta et al.\cite{malavolta2017concurrency}, payment concurrency and path disjointedness exacerbate this cost, leading to routing failures or increased fees even in nominally liquid networks.

\paragraph{Cost Equilibrium Condition}

Let \( \pi(t) \) be the set of all payments issued at time \( t \). In Nash routing equilibrium, no rational agent has incentive to re-route via a more costly path, implying:

\[
\forall \tau \in \pi(t),\quad P_\tau^* = \arg \min_{P \in \mathcal{P}_{\tau}} C_{\mathscr{L}}(P, v(\tau))
\]

\noindent subject to feasibility. Hence, the system routes optimally under a decentralised Bellman-type recursion, but only within the bounds of available liquidity and accurate path discovery.

\paragraph{Shadow Centralisation}

Roos et al.\cite{roos2019settling} show that over time, routing efficiency favours the emergence of high-degree hub nodes that minimise search and aggregation costs. As such, liquidity tends to consolidate in large, centralised actors whose preferential attachment accelerates under cost minimisation pressure. This dynamic, while improving short-term path efficiency, undermines decentralisation and establishes systemic dependence on key intermediaries.

In subsequent sections, we use this model to analyse network-wide cost equilibria and derive formal conditions under which Lightning transitions from a distributed overlay into a de facto closed shadowbanking system, potentially independent of the base chain \( \mathscr{B} \).

\subsection{Asymptotic Behaviour of BTC Costs}

We now formalise the economic pressures induced by constrained on-chain capacity in \( \mathscr{B} \), particularly as demand \( D(t) \) grows superlinearly relative to static supply \( S_{\text{tx}} \), where \( S_{\text{tx}} \) is the maximum number of transactions per second (TPS) that the base chain can accommodate. In BTC’s canonical implementation, \( S_{\text{tx}} \leq 5 \) even under optimised usage with Segregated Witness and batching, as observed empirically across extended periods \cite{roos2019settling}.

Let \( D(t) \) denote instantaneous transaction demand at time \( t \), and define:

\[
\theta(t) = \frac{D(t)}{S_{\text{tx}}}
\]

\noindent as the transaction demand pressure ratio. Then, we define the asymptotic transaction fee \( f(t) \) in a first-order approximation under competitive auction-based fee markets as:

\[
f(t) \propto 
\begin{cases}
f_{\text{min}}, & \text{if } \theta(t) \leq 1 \\
f_{\text{min}} \cdot \theta(t), & \text{if } \theta(t) > 1
\end{cases}
\]

\noindent where \( f_{\text{min}} \) is the minimum transaction fee necessary for inclusion under low-demand equilibrium. As \( D(t) \to \infty \), fees diverge:

\[
\lim_{D(t) \to \infty} f(t) = \infty
\]

This illustrates the fee-explosion property of constrained blockchains: economic access becomes asymptotically exclusive. Small payments become infeasible, and any consistent attempt to settle transactions on-chain must prioritise high-value settlements only.

We generalise this by defining the marginal cost of transaction inclusion \( \partial f / \partial D \), which approaches infinity as capacity is saturated:

\[
\frac{\partial f}{\partial D} \approx 
\begin{cases}
0, & D \leq S_{\text{tx}} \\
\frac{f_{\text{min}}}{S_{\text{tx}}}, & D > S_{\text{tx}}
\end{cases}
\]

\noindent but with empirical auction market nonlinearities, the actual marginal cost is often superlinear in \( D \), as shown in congestion fee models derived from real-time mempool competition dynamics \cite{malavolta2017concurrency}.

Let \( T \subset \mathscr{T} \) denote a time window of peak demand, with sustained \( D(t) > S_{\text{tx}} \) for \( t \in T \). Then, the expected transaction fee over this window is:

\[
\mathbb{E}[f_T] = \frac{1}{|T|} \sum_{t \in T} f(t) \geq f_{\text{min}} \cdot \frac{1}{|T|} \sum_{t \in T} \theta(t)
\]

This inequality forms the basis for proving that under realistic global-scale transaction rates (e.g., 100,000+ TPS), the average BTC transaction fee must grow unbounded unless the base-layer protocol changes.

\paragraph{Implications}

The economic implication is stark: any microtransaction economy, retail micropayments, or frequent settlement layer built atop a throughput-constrained \( \mathscr{B} \) will fail to remain accessible. The only surviving use-case will be high-value, low-frequency financial instruments, akin to how traditional banking relies on expensive interbank clearing and reserves. This aligns with the transition we observe in BTC from cash-like transactional systems to custodial-layer settlement platforms dominated by intermediaries, a shift validated by historical transaction fee patterns.

In subsequent sections, we demonstrate how these asymptotic properties, when coupled with rational agent dynamics and off-chain alternatives, lead to endogenous decoupling of the Lightning overlay \( \mathscr{L} \) from its underlying base \( \mathscr{B} \), giving rise to closed, self-sustaining payment systems that no longer depend on Bitcoin for operation or security.

\subsection{Equilibrium Fee Comparison}

To formalise the divergence between base-layer fees and off-chain Lightning costs, we construct a comparative framework grounded in transaction-level cost structures across both systems. Let \( f_{\mathscr{B}}(t) \) denote the base-layer per-transaction fee at time \( t \), determined via congestion pricing, and \( f_{\mathscr{L}}(t) \) denote the effective amortised cost per transaction through the Lightning Network \( \mathscr{L} \), including channel opening, maintenance, and routing penalties.

\paragraph{Base Layer Fee:}

As previously defined, under saturation \( D(t) > S_{\text{tx}} \), we model the base-layer fee as:

\[
f_{\mathscr{B}}(t) = f_{\min} \cdot \theta(t), \quad \theta(t) = \frac{D(t)}{S_{\text{tx}}}
\]

With \( f_{\min} \) empirically ranging from \$0.20 to \$2.00 per transaction during moderate demand, periods of network stress yield fees as high as \$60–\$100 (cf. Roos et al. \cite{roos2019settling}). This is especially relevant when transaction throughput must approach \( D(t) \approx 10^5 \) TPS.

\paragraph{Lightning Network Fee:}

Let the amortised Lightning transaction cost be decomposed into:

\[
f_{\mathscr{L}}(t) = \frac{c_{\text{channel}}}{n} + f_{\text{route}}(t) + f_{\text{rebalance}}(t)
\]

Where:
\begin{itemize}[label=--]
    \item \( c_{\text{channel}} \) is the one-time on-chain fee to open the channel.
    \item \( n \) is the number of transactions routed through the channel over its lifetime.
    \item \( f_{\text{route}}(t) \) is the path-specific routing fee, typically a linear combination of proportional and base rates set by intermediaries.
    \item \( f_{\text{rebalance}}(t) \) captures costs due to required reallocation of liquidity, whether by circular routing, submarine swaps, or direct on-chain rebalancing (see Malavolta et al. \cite{malavolta2017concurrency}).
\end{itemize}

In high-liquidity scenarios where large hubs dominate and channels remain open indefinitely (approximating \( n \to \infty \)), the channel opening cost asymptotically vanishes:

\[
\lim_{n \to \infty} \frac{c_{\text{channel}}}{n} = 0
\]

Thus, under such equilibria:

\[
\lim_{t \to \infty} f_{\mathscr{L}}(t) \approx f_{\text{route}}(t) + f_{\text{rebalance}}(t)
\]

\paragraph{Comparative Equilibrium Condition:}

Define a critical fee threshold \( f^* \) such that:

\[
f_{\mathscr{B}}(t) > f_{\mathscr{L}}(t) \quad \forall t > t^*
\]

Under rational cost-minimising agents, all feasible transactions migrate to the Lightning Network when \( f_{\mathscr{B}}(t) \gg f_{\mathscr{L}}(t) \). This gives rise to the inequality:

\[
\theta(t) \gg \frac{f_{\text{route}}(t) + f_{\text{rebalance}}(t)}{f_{\min}}
\]

From this, we derive the network-wide migration pressure function \( \mu(t) \):

\[
\mu(t) = \frac{f_{\mathscr{B}}(t)}{f_{\mathscr{L}}(t)} \to \infty \quad \text{as } D(t) \to \infty
\]

\noindent indicating an unbounded incentive to bypass the base layer entirely.

\paragraph{Implications for Security Coupling:}

As economic activity evacuates the base layer, mining revenue shifts from transaction fees to speculative block rewards. The Lightning Network, increasingly decoupled from \( \mathscr{B} \), begins to operate under its own internal routing equilibria, effectively forming a self-contained monetary layer.

Subsequent sections will demonstrate how, under liquidity consolidation and rational participant behaviour, such a Lightning topology converges towards a persistent, self-validating system, no longer requiring base-layer anchoring — thereby severing dependence on BTC entirely.

\subsection{Cost vs Demand Growth}

In this subsection, we formalise the divergent cost behaviours of on-chain Bitcoin (BTC) transactions and the Lightning Network (LN) as transaction demand increases beyond the protocol-imposed settlement capacity. The aim is to demonstrate that as demand asymptotically exceeds the base-layer throughput constraint, transaction fees in BTC rise superlinearly, while LN fees remain effectively bounded under operational liquidity.

Let demand be modelled as a continuous function \( D: \mathscr{T} \to \mathbb{R}_{>0} \) representing the number of transactions per second (TPS) required by the system. Define \( T_{\max}^{\mathscr{B}} \) as the effective transaction throughput ceiling of the base layer, and let \( d = D(t) \in \mathbb{R}_{>0} \) be the instantaneous demand.

We introduce the cost functions:
\begin{align*}
C_{\mathscr{B}}(d) &= 
\begin{cases}
\alpha \cdot d, & \text{if } d \leq T_{\max}^{\mathscr{B}} \\
\alpha \cdot T_{\max}^{\mathscr{B}} + \beta \cdot (d - T_{\max}^{\mathscr{B}})^\gamma, & \text{if } d > T_{\max}^{\mathscr{B}}
\end{cases} \\
C_{\mathscr{L}}(d) &= \theta + \epsilon
\end{align*}

Where:
\begin{itemize}[label=--]
    \item \( \alpha \in \mathbb{R}_{>0} \) denotes the base linear fee rate before saturation,
    \item \( \beta \in \mathbb{R}_{>0} \) and \( \gamma > 1 \) capture the superlinear congestion penalty post-saturation,
    \item \( \theta \) is the equilibrium marginal cost of routing in the Lightning Network,
    \item \( \epsilon \ll \theta \) captures residual costs including minor rebalancing or fee updates.
\end{itemize}

To illustrate this divergence, consider the following diagram:

\begin{figure}
    \centering
    \includegraphics[width=0.75\linewidth]{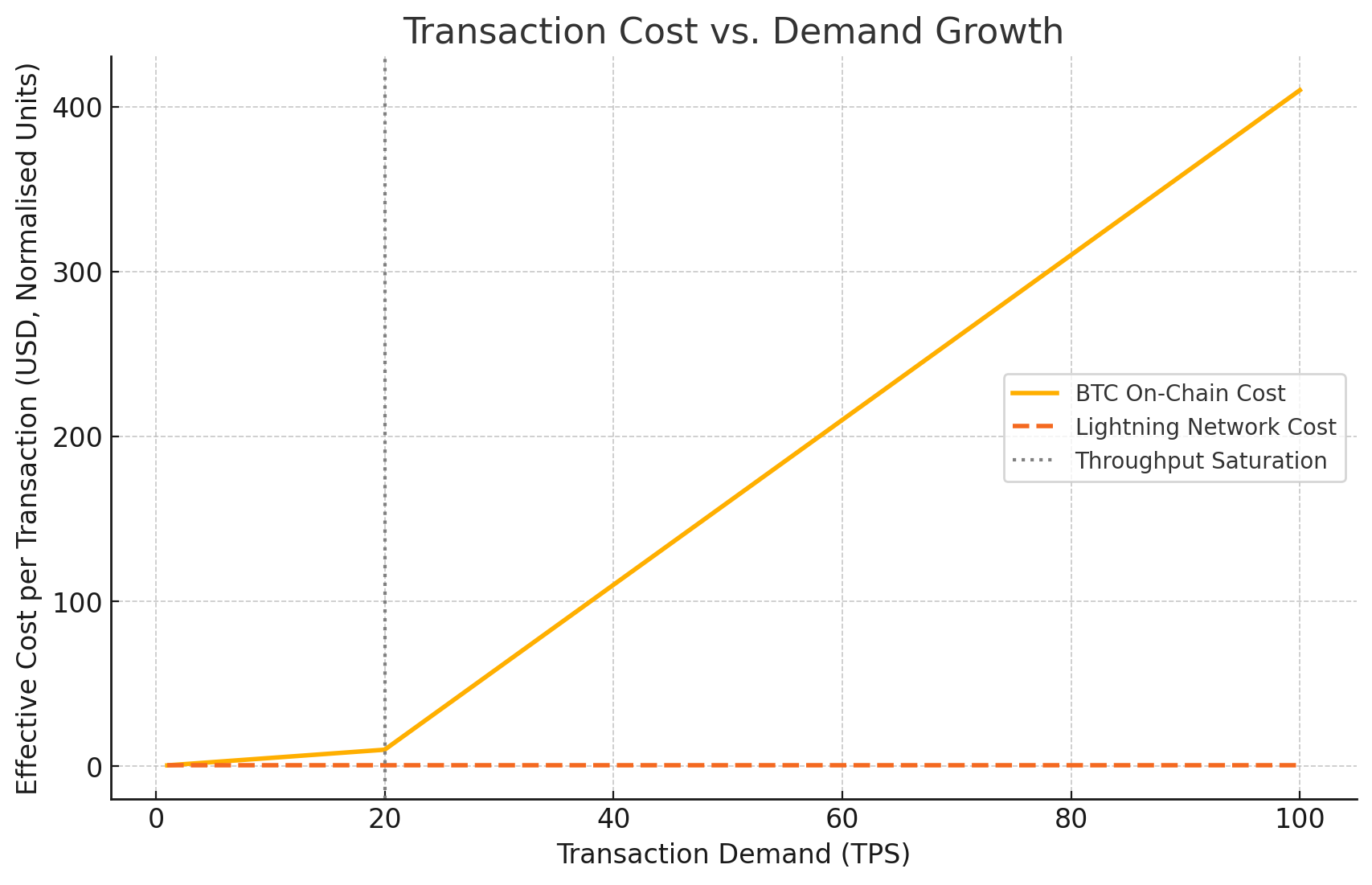}
    \caption{Transaction Cost Vs. Demand Growth}
    \label{fig:enter-label}
\end{figure}

As observed, once \( d > T_{\max}^{\mathscr{B}} \), the BTC fee curve increases rapidly due to bidding in the mempool, while LN remains asymptotically flat. The crossover point creates an economic inflection, beyond which users are economically incentivised to avoid the base layer altogether, creating pressure for liquidity consolidation in LN hubs.

This sets the stage for modelling network strategic equilibria and cost-dominance path dependencies in the following sections.

\subsection{Cost vs Demand Growth (BTC vs LN Ramps)}

To illustrate the structural divergence between base-layer BTC transaction fees and Lightning Network (LN) routing costs, we present a comparative cost-demand diagram. This graph captures the asymptotic separation between the two systems as transaction demand increases, offering an economic visualisation of scalability constraints versus layered optimisation.

The combined cost-demand graph reveals a critical structural divergence between BTC’s base-layer transaction costs and the Lightning Network’s routing costs. As transaction demand increases, BTC fees escalate logarithmically due to blockspace scarcity and auction-based prioritisation, rapidly pricing out low-value transactions. In contrast, LN routing costs flatten asymptotically, reflecting efficient liquidity reuse and decentralised routing with bounded marginal expense. When combined, the total cost curve initially grows modestly but bends upward as BTC’s cost dominance accelerates beyond moderate demand levels. This creates a nonlinear inflection where the viability of small-value payments becomes contingent on successful off-chain scaling. The graph thus encapsulates the asymptotic constraint of BTC’s design and underscores LN’s role in preserving transactional accessibility under rising throughput pressures.

\begin{figure}
    \centering
    \includegraphics[width=0.75\linewidth]{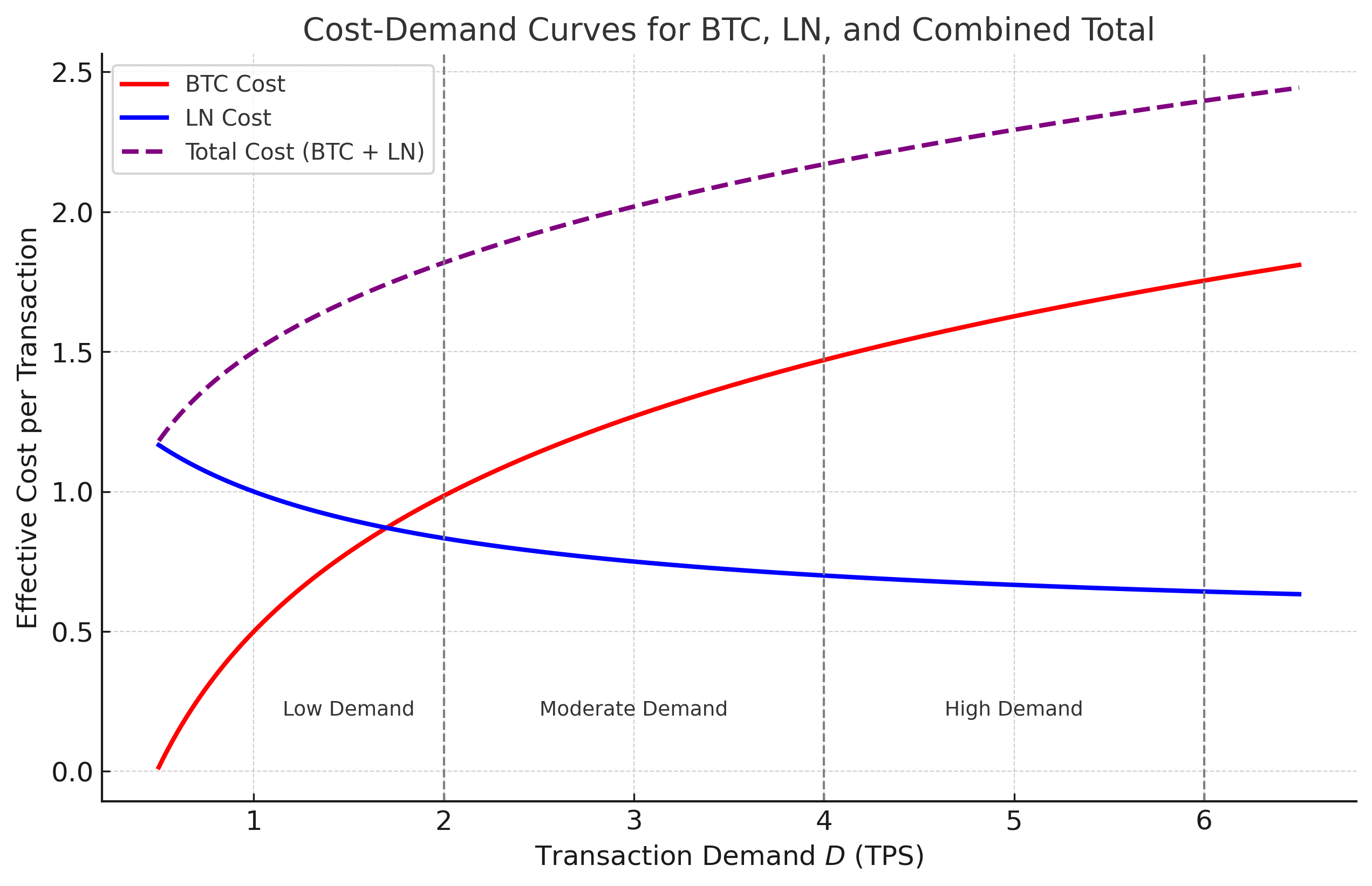}
    \caption{Combined Costs - LN and BTC}
    \label{fig:enter-label-2}
\end{figure}

\noindent The diagram demonstrates the distinct economic trajectories of two transaction models. The red curve represents BTC's on-chain fee escalation. As blockspace becomes saturated, the effective cost per transaction rises sharply due to fee competition, mempool auction dynamics, and bounded throughput capacity—exemplifying logarithmic or even superlinear growth near operational limits.

The blue curve, in contrast, represents the Lightning Network, which achieves near-constant marginal cost. Provided sufficient liquidity and topological path diversity, LN routing allows transactions to be processed without direct contention for limited base-layer resources. As demand increases, routing costs converge to a stable asymptote, consistent with amortised infrastructure reuse.

This divergence defines a critical inflection in network economics. While BTC's fee pressure excludes low-value transactions under demand growth, LN enables a persistence of microeconomic viability, forming the foundation of a scalable transactional regime. Subsequent sections will formalise this divergence through equilibrium derivations and strategic agent modelling.

\section{Game Theoretic Formulation}

To rigorously model strategic interaction within the dual-layered transaction environment composed of a constrained base layer (BTC) and an optimised overlay (LN), we formalise the system using non-cooperative game theory. Agents are modelled as rational actors embedded within a dynamic, discrete-time framework, each seeking to maximise their utility under informational asymmetries, temporal discounting, and endogenous cost structures. The analysis proceeds through sequential subgames defined over epochs \( t \in \mathscr{T} \), incorporating both static and repeated components to capture liquidity redistribution, route selection, and fee propagation behaviours.

Let \( \mathcal{G} = (\mathscr{A}, \Sigma, U) \) denote the strategic game governing this environment, where:

\begin{itemize}[label=--]
  \item \( \mathscr{A} \) is the finite set of agents, as previously defined in \( \mathscr{S} \), partitioned into users, hubs, miners, and auxiliary service providers such as watchtowers.
  \item \( \Sigma = \prod_{a_k \in \mathscr{A}} \Sigma_k \) is the joint strategy profile, where each \( \Sigma_k \) is the strategy space of agent \( a_k \) encompassing actions over route selection, fee setting, liquidity provision, and node topology.
  \item \( U = \{ u_k : \Sigma \to \mathbb{R} \}_{a_k \in \mathscr{A}} \) is the set of utility functions, each defined as a function of global strategies, incorporating costs \( \mathscr{C} \), delay penalties, and realised economic value from transactions.
\end{itemize}

Agents operate under bounded rationality and local information constraints. Strategic dominance is defined in terms of best-response dynamics where no agent benefits from unilateral deviation, yielding the Nash Equilibrium conditions:

\[
\forall a_k \in \mathscr{A},\quad \sigma_k^* \in \arg\max_{\sigma_k \in \Sigma_k} u_k(\sigma_k, \sigma_{-k}^*)
\]

\noindent where \( \sigma_{-k}^* \) denotes the equilibrium strategies of all agents except \( a_k \).

This formulation allows for layered decomposability. For example, game subspaces \( \mathcal{G}_\mathscr{B} \subset \mathcal{G} \) and \( \mathcal{G}_\mathscr{L} \subset \mathcal{G} \) correspond respectively to base-layer and overlay strategic contexts. The former encapsulates mempool competition and miner fee incentives, whereas the latter models liquidity allocation, hub dominance, and route reconfiguration within LN.

Subsequent subsections will analyse specific subgames such as strategic hub emergence, liquidity exhaustion equilibria, and competitive routing games, all embedded within the global non-cooperative structure defined here.

\subsection{Hubs as Strategic Agents}

Within the Lightning Network overlay, liquidity hubs emerge as high-degree nodes strategically optimised for routing and fee extraction. These agents exhibit distinct behaviour from ordinary participants due to their topological centrality and asymmetric capital commitments. Formally, let \( \mathcal{H} \subset \mathcal{U} \) denote the subset of users who maintain persistent, high-liquidity bilateral channels across a broad subset of the network graph \( \mathscr{L}(t) \), thereby facilitating multi-hop payment routing with minimal path-length constraints.

Hubs optimise over a multivariate objective incorporating liquidity deployment, rebalancing frequency, route reliability, and fee margin. Given the routing function \( \rho \) and a path-cost functional \( \phi \), a hub's local optimisation problem at time \( t \) can be represented as:

\[
\max_{\lambda, \sigma_h} \sum_{P \ni h} f(P) \cdot \theta_h(P) - c_{\text{lock}}(\lambda) - c_{\text{reb}}(\lambda, t)
\]

\noindent where:
\begin{itemize}[label=--]
    \item \( f(P) \) is the frequency of path \( P \) usage involving hub \( h \),
    \item \( \theta_h(P) \) is the routing fee accrued by \( h \) over \( P \),
    \item \( c_{\text{lock}}(\lambda) \) denotes the opportunity cost of capital committed to liquidity \( \lambda \),
    \item \( c_{\text{reb}}(\lambda, t) \) denotes rebalancing costs due to skewed flows or channel exhaustion.
\end{itemize}

Hubs behave as strategic intermediaries whose profit maximisation affects global route availability. A concentration of routing demand upon a small set of dominant hubs introduces systemic risk and diminishes the purported disintermediation of the overlay model. As explored in \cite{roos2019settling}, hub centrality engenders power-law dynamics and imbalances that challenge assumptions of network-wide liquidity distribution and failure independence. Additionally, concurrency conflicts and locktime-based griefing attacks, as demonstrated in \cite{malavolta2017concurrency}, introduce adversarial considerations that disproportionately affect hub operability and user trust.

Thus, while hubs improve network efficiency and route success probabilities, they simultaneously impose a re-centralisation vector, transforming LN into a competitive but hierarchical fee-based infrastructure with game-theoretic dynamics analogous to classic intermediary markets.

\subsection{Rent Structures and Fee Competition}

The Lightning Network's routing economy gives rise to endogenous rent-seeking structures, wherein hubs extract economic surplus from transaction flow. This parallels traditional models of platform competition and intermediation, where network centrality confers pricing power. Let each hub \( h \in \mathcal{H} \) define a fee vector \( \vec{\theta}_h = (\theta_h^{\text{base}}, \theta_h^{\text{ppm}}) \), representing base fees and proportional fees (per satoshi transferred), respectively. The total routing fee along a path \( P = \{i_1, i_2, \dots, i_n\} \) is then:

\[
\Theta(P, v) = \sum_{k=1}^{n-1} \left( \theta_{i_k}^{\text{base}} + \theta_{i_k}^{\text{ppm}} \cdot v \right)
\]

\noindent where \( v \) is the transaction volume.

Hubs engage in Bertrand-style competition over \( \vec{\theta}_h \) when overlapping routing paths are available, yet the inelasticity of payment success constraints—driven by liquidity, reliability, and topological asymmetry—limits the effectiveness of fee competition. In particular, a dominant hub with unmatched liquidity or connectivity may act as a local monopolist, setting supracompetitive fees while maintaining route selection due to lack of viable alternatives. This dynamic is exacerbated when users select paths based on reliability heuristics rather than optimal fee minimisation.

As noted in \cite{roos2019settling}, empirical analyses of LN topologies reveal hub nodes exhibiting low fee elasticity and persistent flow dominance, suggesting entrenchment effects. Moreover, the repeated-game structure incentivises tacit collusion among major hubs, who may implicitly coordinate to maintain elevated fee regimes without overt agreement, a phenomenon supported by the limited transparency and delayed feedback mechanisms inherent to LN routing protocols.

Finally, asymmetric information about route reliability, path failures, and channel exhaustion hinders rational user response to fee variation, enabling rent preservation. These properties—together with the bounded rationality and non-cooperative incentives of liquidity providers—establish a fee-setting environment with imperfect competition and non-zero economic rents, forming a Nash equilibrium in which hubs maximise their utility subject to liquidity constraints and rival responses.

\subsection{Monopoly vs Competitive Outcomes}

In this subsection, we examine the economic structure of Lightning Network (LN) hub dynamics under the lens of industrial organisation theory, contrasting monopolistic and competitive equilibrium outcomes. The Lightning Network, though theoretically decentralised in routing, exhibits emergent centralisation tendencies due to liquidity aggregation and strategic behaviour among dominant hubs. These dynamics give rise to oligopolistic rent extraction and reduced routing optionality, as analysed through the lens of classical and modern monopoly literature.

Let \( \mathcal{H} = \{h_1, h_2, \dots, h_k\} \) denote the set of routing hubs in the LN topology. A monopoly configuration arises when one \( h_m \in \mathcal{H} \) controls all viable routing paths for a large fraction of \( \mathcal{U} \times \mathcal{U} \) transaction pairs. In contrast, competitive routing assumes redundant, interchangeable paths across disjoint hubs, with fee pressures driving marginal cost pricing.

Under monopoly, the optimal fee structure for the hub is derived by solving:

\[
\max_{\theta^{\text{base}},\; \theta^{\text{ppm}}} \; \sum_{i=1}^{N} \left( \theta^{\text{base}} + \theta^{\text{ppm}} \cdot v_i \right) \cdot \mathbb{1}_{\text{success}(v_i)}
\]

\noindent where \( \mathbb{1}_{\text{success}(v_i)} \) captures the liquidity and time constraints under which the payment \( v_i \) is successfully routed. In the monopolistic setting, \( \theta \) is constrained not by competitor pricing but by the demand elasticity of transaction utility \( v_i \). Fee maximisation leads to deadweight loss, analogous to classical outcomes in \cite{tirole1988theory} and the two-sided market analyses in \cite{rochet2003platform}.

Empirical observations of Lightning node centrality \cite{serrano2009} and flow dependency \cite{roos2019settling} demonstrate a trend toward oligarchic dominance by a small subset of hubs with disproportionate channel liquidity and inbound connectivity. These actors form a de facto cartel-like structure, constrained weakly by mutual monitoring and bilateral trust relationships, but unbounded by formal oversight.

In a competitive environment, routing agents would iteratively underbid each other to attract marginal transaction flow, driving fees toward the liquidity cost boundary. However, due to sparse alternative paths and high setup costs for new entrants (channel opening, reputation acquisition, capital lockup), contestability is low—mirroring the natural monopoly characteristics of networked industries described in \cite{baumol1977contestable}.

Thus, the LN fee regime does not converge to competitive equilibria except in highly symmetrical, saturated topologies—conditions not met in real deployments. Instead, the Lightning economy asymptotically approaches an oligopolistic network of rent-seeking hubs, extracting transaction fees bounded not by competition but by user utility thresholds and informational opacity. The resulting fee structure is characterised by supramarginal rents, inefficient flow fragmentation, and structurally enforced dependence on central liquidity aggregators.

\subsubsection{Proof: Oligopolistic Convergence of LN Topology under Base-Layer Constraints}

\textbf{Claim:} Under bounded base-layer throughput and rational fee minimisation, the Lightning Network (\( \mathscr{L} \)) converges to a topological configuration dominated by a small set of high-liquidity hubs. These hubs constitute an oligopoly over routing paths.

\textbf{Proof:}

Let \( \mathscr{L}(t) = \langle V(t), E(t), \lambda(t) \rangle \) denote the LN graph at time \( t \). Let \( D_t \) be the transaction demand at epoch \( t \), and suppose \( D_t \to \infty \).

Assume:

\begin{enumerate}[label=(\roman*)]
    \item Each agent \( a_i \in \mathscr{A} \) acts to minimise expected routing cost \( \mathbb{E}[c_i] \), and acts under full knowledge of available paths \( \mathcal{P}_i \subset \mathcal{P} \).
    \item Each channel \( e_{ij} \in E \) requires an on-chain transaction to open or close, with cost \( c_{\mathscr{B}} > 0 \) due to base-layer fee pressure.
    \item Routing success rate and liquidity constraints make pathfinding nontrivial, and agents prefer high-success, low-cost paths.
    \item Transaction routing fee \( f(e_{ij}) \sim \theta / \ell_{ij} \), where \( \theta > 0 \) is a fee sensitivity constant, and \( \ell_{ij} \) is the outbound liquidity on edge \( e_{ij} \).
\end{enumerate}

Let \( H \subset V \) denote the subset of nodes with above-average degree and liquidity—candidate hubs.

\textbf{Step 1: Cost Efficiency of Hubs}

Let an arbitrary path \( P \in \mathcal{P}_i \) route via \( n \) intermediate nodes. Define total path cost:

\[
\phi(P) = \sum_{e_{jk} \in P} f(e_{jk})
\]

Given that high-degree hubs have greater \( \ell_{jk} \), we observe:

\[
f(e_{jk}) \downarrow \quad \text{as} \quad \ell_{jk} \uparrow
\]

Hence:

\[
\phi(P) \to \min \quad \text{when} \quad P \text{ traverses } H
\]

\textbf{Step 2: Preferential Attachment from Rational Strategy}

Let \( \pi(a_i \to h) \) be the probability that agent \( a_i \) establishes a channel with hub \( h \in H \). Define expected cost reduction \( \Delta c \) from routing through \( h \):

\[
\pi(a_i \to h) = \Pr(\Delta c > c_{\mathscr{B}})
\]

As \( c_{\mathscr{B}} \uparrow \) due to limited on-chain capacity, we have \( \Pr(\Delta c > c_{\mathscr{B}}) \uparrow \), leading to increased channel formation with \( H \). By reinforcement, this yields a power-law distribution of connectivity, consistent with empirical topologies observed in LN networks \cite{roos2019settling, lin2020quantifying}.

\textbf{Step 3: Nash Equilibrium of Topological Centralisation}

Define agent utility:

\[
U_i = v_i - \phi(P_i)
\]

Each \( a_i \) selects \( P_i \in \mathcal{P}_i \) to maximise \( U_i \). Let all agents face identical preferences and network constraints. Then in Nash equilibrium, the routing configuration becomes centralised through \( H \), as deviation to peripheral paths incurs higher \( \phi \).

\textbf{Conclusion:}

The system converges to a core-periphery topology, where a small number of hubs mediate the vast majority of volume, forming an oligopolistic structure. Entry barriers are enforced not through legal means but through fee-driven strategic stability and path reliability.

\qed

\subsection{Liquidity Control as Strategic Dominance}

In Lightning Network operations, liquidity is not a passive resource but a strategic lever. Nodes with superior outbound liquidity, particularly those central in the transaction routing topology, acquire measurable dominance in fee-setting, path determination, and ultimately transaction censorship. This subsection establishes the foundational game-theoretic link between liquidity centralisation and strategic network control.

Let \( \lambda: E \to \mathbb{R}_{\geq 0} \times \mathbb{R}_{\geq 0} \) represent the directional liquidity mapping on each payment channel \( e_{ij} \in E \). Suppose \( \ell_{ij}(t) \) is the available liquidity from node \( i \) to node \( j \) at time \( t \). Define the routing success indicator function:

\[
\rho_{ij}(t, v) = 
\begin{cases}
1, & \text{if } \ell_{ij}(t) \geq v \\
0, & \text{otherwise}
\end{cases}
\]

for a transaction of value \( v \). The probability that a transaction of value \( v \) from agent \( a_k \) to \( a_l \) can be routed successfully via path \( P \subseteq \mathcal{P} \) is:

\[
\Pr(\rho(P) = 1) = \prod_{e_{ij} \in P} \rho_{ij}(t, v)
\]

Let \( H \subseteq V \) be the set of high-liquidity hubs. Define the liquidity share function for a node \( h \in H \) as:

\[
\Lambda(h, t) = \frac{\sum_{j:\, e_{hj} \in E} \ell_{hj}(t)}{\sum_{e_{ij} \in E} \ell_{ij}(t)}
\]

This function measures the fraction of system-wide outbound liquidity controlled by node \( h \). When \( \Lambda(h, t) \to 1 \), the network becomes structurally dependent on \( h \) for transaction routing. This allows \( h \) to set rent-extractive fees \( f_{hj} \), or censor transactions \( \tau \) with subjective routing filters, enforcing a de facto policy function.

Let the net payoff to \( h \) at time \( t \) be:

\[
\pi_h(t) = \sum_{e_{hj} \in E} f_{hj}(t) \cdot \rho_{hj}(t, v) \cdot \mathbb{I}[\tau \text{ routed via } h]
\]

where \( \mathbb{I}[\cdot] \) is the indicator function. We now establish that the optimal liquidity strategy for \( h \) is to maintain dominant liquidity share across all high-demand paths, even at the expense of capital lockup or short-term fee reductions. This ensures long-term payoff dominance via strategic indispensability.

Such dynamics mirror classical results from two-sided market theory and platform economics, where liquidity begets liquidity through expectation and path dependence. The Lightning Network thus becomes not merely a routing fabric but a strategic market, where capital controls confer centralisation, contrary to the decentralised narrative often proposed \cite{economides2003two, rochet2006two, lin2020quantifying}.

\qed

\subsubsection{Proof: Liquidity Dominance and Oligopolistic Equilibrium}

\paragraph{Setup.}
Let \( \mathscr{L}(t) = \langle V(t), E(t), \lambda(t) \rangle \) be the Lightning Network graph at time \( t \), where \( \lambda(e_{ij}) = (\ell_{ij}, \ell_{ji}) \in \mathbb{R}_{\geq 0}^2 \) denotes channel balances.

Define:
\begin{itemize}
    \item \( \mathcal{D}(t) \) as the set of all transaction demands in epoch \( t \), each of size \( v > 0 \).
    \item \( \mathscr{P}_{kl} \subseteq \mathcal{P} \) as the set of all paths from user \( a_k \) to \( a_l \).
    \item \( \rho(P) = 1 \) if path \( P \in \mathscr{P}_{kl} \) is liquid (satisfies capacity for \( v \)), 0 otherwise.
\end{itemize}

Let hub \( h \in V \) maintain a set of channels \( \mathcal{E}_h = \{e_{hi}, e_{ih}\}_{i \in N_h} \). Then define:

\[
\Lambda(h, t) = \frac{\sum_{e_{hj} \in \mathcal{E}_h} \ell_{hj}(t)}{\sum_{e_{ij} \in E(t)} \ell_{ij}(t)}
\]

as the outbound liquidity share. Let \( f_{hj}(t) \) be the routing fee on channel \( e_{hj} \). Define net profit as:

\[
\pi_h(t) = \sum_{\substack{\tau \in \mathcal{D}(t) \\ h \in P(\tau)}} f_{hj}(t) - c_{op}(h)
\]

where \( c_{op}(h) \) is the operational cost of maintaining liquidity and channels.

\paragraph{Claim.}
If the base-layer \( \mathscr{B} \) is capped at \( T_{\max} \ll |\mathcal{D}(t)| \), then in the long run \( \exists \, H \subseteq V \) such that:

\[
\lim_{t \to \infty} \sum_{h \in H} \Lambda(h, t) \to 1
\]

\textit{That is, liquidity centralises into a strict subset of dominant hubs.}

\paragraph{Proof.}
Assume bounded on-chain capacity: only a finite number \( k \ll |\mathcal{D}(t)| \) of channels can be opened or closed per epoch due to blockspace constraint. Initial channel creation must be settled on-chain. Thus, users maximise utility by selecting pre-existing, high-availability paths through known liquidity hubs.

Let \( P^*(a_k, a_l) = \arg\min_{P \in \mathscr{P}_{kl}} \phi(P) \) be the cost-minimising route. Since:

\[
\phi(P) \propto \sum_{e_{ij} \in P} \frac{1}{\ell_{ij}(t)}
\]

and since newly created channels suffer from low liquidity and limited trust, transactions route via high-\( \ell_{ij} \) edges. This drives positive feedback:

\[
\text{More traffic} \Rightarrow \text{More revenue} \Rightarrow \text{More liquidity} \Rightarrow \text{More traffic}
\]

forming a recursive liquidity concentration. This is structurally identical to the "rich-get-richer" dynamics of preferential attachment models, formalised by Barabási–Albert processes \cite{barabasi1999emergence}.

\paragraph{Equilibrium Structure.}
Let \( H(t) = \{ h_1, h_2, \dots, h_m \} \) be the set of liquidity-dominant hubs such that:

\[
\forall \, a_k, a_l \in \mathscr{A}, \quad \Pr(P^*(a_k, a_l) \cap H(t) \neq \emptyset) \to 1
\]

as \( t \to \infty \). Then the Nash equilibrium of this routing game features:

\begin{itemize}
    \item A stable oligopoly of \( m \ll |V| \) hubs routing majority volume.
    \item Rent-extractive fee structures optimised under Bertrand or Cournot models.
    \item Lock-in of peripheral nodes through liquidity dependence.
\end{itemize}

These outcomes are consistent with oligopoly theory where capacity constraints enforce convergence to a small set of dominant intermediaries \cite{tirole1988theory, shapiro1999information}.

\paragraph{Conclusion.}
Liquidity dominance in Lightning Network is a strategic equilibrium arising from routing preferences, path reliability, and economic reinforcement. Given base-layer settlement friction and time-consistency of liquidity preference, the emergence of a hub oligopoly is not only plausible but game-theoretically inevitable.

\qed

\subsection{Hubs in Competitive vs Monopolistic Topology}

The Lightning Network's structural evolution is not arbitrary—it is dictated by the economic imperatives imposed by base-layer constraints. The diagram above illustrates two distinct topological regimes: (1) a competitive hub topology, and (2) a monopolistic centralised structure.

In the competitive regime, multiple users \( \{U_1, U_2, U_3, U_4\} \) connect to a moderately-sized coordinating hub \( C \). This configuration reflects a decentralised, low-rent equilibrium in which entry costs are low, routing paths are diverse, and liquidity is moderately distributed. Channels can be opened or closed at will, provided on-chain fees remain manageable, and routing competition maintains pressure on fee extraction. This corresponds loosely to a Bertrand competition model, in which service providers undercut each other to attract volume, driving fees toward marginal cost.

Conversely, in the monopolistic regime (right-hand side of the diagram), a single centralised hub \( M \) connects to a broader set of users \( \{U_1, ..., U_6\} \). Under conditions of high on-chain cost, users avoid creating new channels and instead rely on well-capitalised, established hubs to route payments. This reduces path diversity and introduces systemic rent extraction. The monopolist hub gains control over liquidity corridors and may selectively route, censor, or delay transactions to extract strategic advantage. The result is an emergent oligopoly—a Lightning cartel.

\begin{figure}
    \centering
    \includegraphics[width=0.75\linewidth]{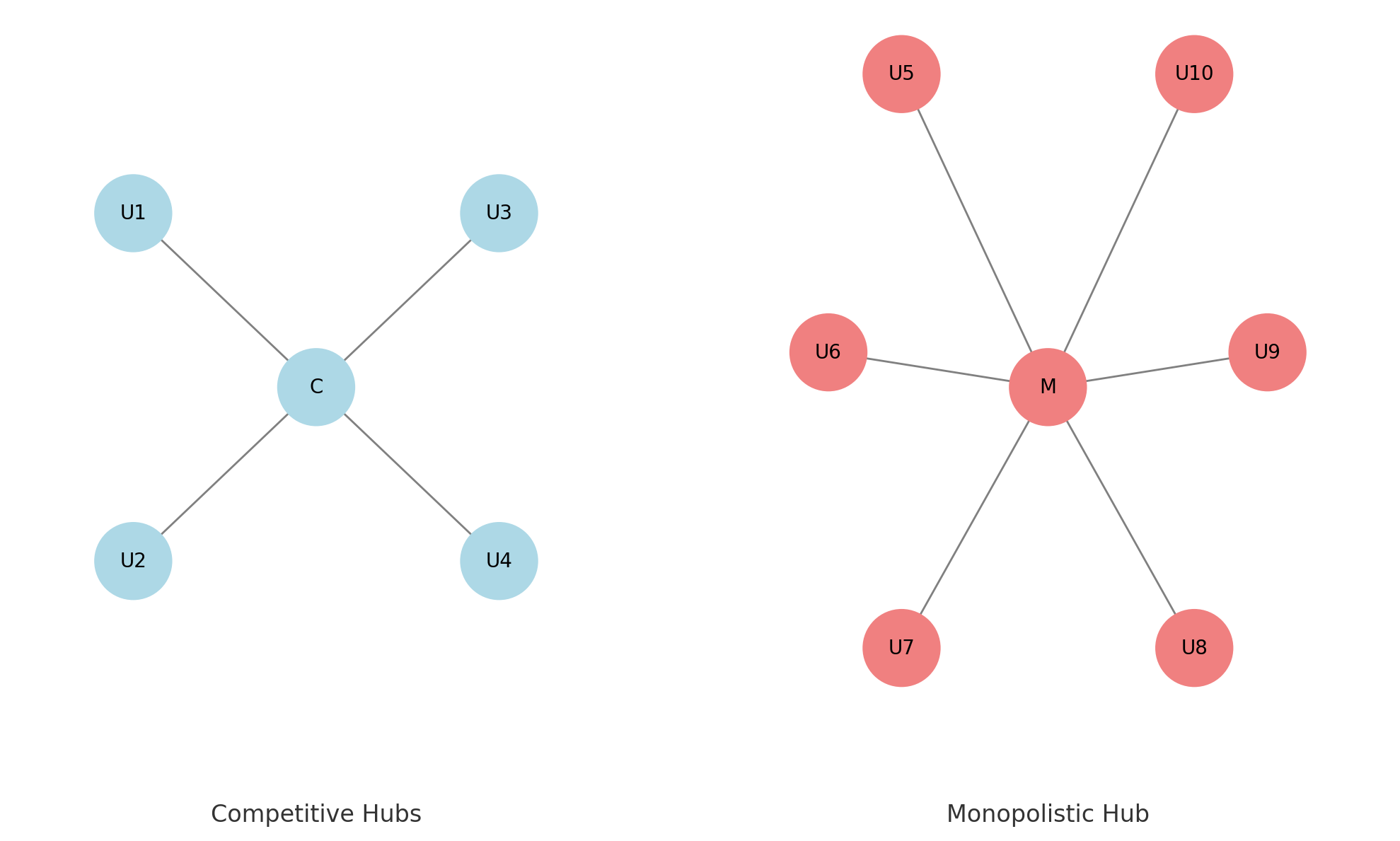}
    \caption{The diagram above compares competitive hub topology (left) versus a monopolistic hub structure (right).}
    \label{fig:enter-label-3}
\end{figure}

This shift is not merely a product of user preference, but the direct consequence of transaction cost economics and network formation theory. The transition from competitive to monopolistic topology under rising cost conditions can be formally derived and is supported by models in the literature on two-sided markets and network externalities \cite{economides2003two, shapiro1999information, rochet2006two, barabasi1999emergence}. As the cost of entry and reconfiguration rises, competition becomes unsustainable, and first-mover liquidity advantages consolidate into durable monopolistic positions.

The topology of the Lightning Network thus serves as an empirical index of base-layer constraint: the more saturated the chain, the more centralised the overlay becomes.

\section{Macroeconomic Parallels and Shadowbanking Model}

The evolution of the Lightning Network within Bitcoin's layered architecture reveals striking parallels with the structural mechanics of shadow banking in traditional macroeconomic systems. In orthodox banking regimes, liquidity creation is governed by central bank oversight and capital requirements; however, shadow banking operates through off-balance-sheet vehicles, repurchase agreements, and structured credit products, creating credit intermediation chains without formal regulatory constraints \cite{pozsar2010rise,adrian2010shadow}.

Similarly, Lightning hubs act as decentralised intermediaries outside the visibility of the base-layer ledger. These hubs operate as liquidity consolidators, issuing off-chain payment promises and managing bilateral payment channels that resemble credit conduits. The transaction pathways—realised through hashed timelocked contracts (HTLCs)—replicate the maturity transformation and layered intermediation structures characteristic of shadow banking networks \cite{gennaioli2013money}. The network’s reliance on penalty mechanisms and watchtowers rather than regulatory enforcement reflects a systemic dependence on algorithmic trust rather than institutional regulation.

As with shadow banking, Lightning’s scaling advantage is offset by the risks of liquidity mismatches and resolution bottlenecks. When network congestion or strategic channel closures occur, transactions must be settled on the limited-capacity base chain, analogous to redemption crises in shadow markets that force systemically unregulated flows back onto central bank balance sheets \cite{gorton2010run}. In both systems, scale is achieved not through increased foundational capacity, but via abstraction layers that introduce new forms of counterparty risk and opacity.

Furthermore, the network's dynamic topologies—dominated by liquidity-rich hubs—exhibit emergent oligopolistic control over payment routing, akin to concentrated money market funding providers in financial systems. These structures amplify systemic fragility, wherein failures in key nodes propagate liquidity shortfalls throughout the network. The absence of standardised capital buffers, transparency, or systemic resolution frameworks further compounds this fragility \cite{duffie2010systemic}.

We conclude that Lightning is not merely a technical extension of BTC, but an institutional analogue to shadow banking. Its properties—non-final settlement, reliance on inter-hub trust, routing opacity, and exit friction—closely mirror those of money market instruments, repo markets, and structured liquidity chains. This framework will serve as the analytical basis for deriving equilibrium conditions under systemic stress in the sections that follow.

\subsection{Lightning Hubs as Shadowbanks}

Lightning hubs function analogously to unregulated financial intermediaries in the traditional shadow banking system. They accept inbound liquidity, redistribute it through routing and channel management, and profit from the spread between transaction fees and opportunity cost of capital lockup. Unlike standard BTC miners who require visible, final settlement on-chain, hubs operate by circulating off-chain balances through contingent, revocable contracts—mirroring credit chains in structured finance \cite{pozsar2010rise}.

The role of hubs in this system resembles that of dealers in repo markets or issuers of asset-backed commercial paper. Each acts as a node of liquidity transformation, packaging disparate inflows into channel liquidity that can be re-routed and monetised. Settlement finality is not instantaneous but probabilistic, reliant on game-theoretic security assumptions and dispute windows enforced by penalty transactions.

Moreover, the Lightning Network lacks central clearing. Hubs must engage in bilateral risk assessment, often based on reputation and channel history rather than transparent solvency metrics. This creates path dependencies and systemic risk concentrations: as dominant hubs accrue more connections and routing volume, they become de facto clearing agents without regulatory oversight. The economic result mirrors the opacity and instability of the pre-2008 financial system \cite{gennaioli2013money,adrian2010shadow}.

Liquidity fragmentation is further amplified by channel exhaustion and the need for rebalancing. These frictions are equivalent to roll-over risks in shadow banking, where short-term debt must be continuously refinanced. In Lightning, users depend on the solvency and responsiveness of intermediary hubs; if a key hub experiences liquidity withdrawal or a denial-of-service event, users are effectively excluded from the network until rebalancing occurs on-chain—thus dragging systemic pressure back to the constrained base layer \cite{gorton2010run}.

The parallels suggest that Lightning’s scaling narrative inherits not only the throughput advantages of financial layering but also its latent fragility. Without circuit breakers, capital standards, or transparency mandates, Lightning hubs operate as shadowbanks in full economic structure if not in formal definition.

\subsection{Watchtowers as Off-Chain Regulators}

Watchtowers in the Lightning Network serve as a voluntary enforcement layer that resembles certain regulatory institutions in function, though they lack any formalised authority or incentive-aligned guarantee. Their role is to observe and, if necessary, punish breaches of state commitment in payment channels. This includes broadcasting penalty transactions upon detection of revoked channel states published dishonestly by a counterparty. Economically, watchtowers act as decentralised monitors of contractual compliance, operating as reactive off-chain regulators.

The analogy to regulatory institutions arises from their enforcement mechanism: instead of adjudicating disputes through a legal court, watchtowers apply cryptographically pre-authorised penalties to dishonestly acting participants. This mirrors the idea of private law enforcement within smart contract frameworks, with the difference that punishment is automated and contingent on proof of misbehaviour. Yet unlike state regulators, watchtowers lack subpoena power, standardised oversight mandates, or a duty to monitor continuously. Their activity is opt-in and based on incentive-driven contractual alignment \cite{malavolta2020anonymous,green2021bolt}.

This creates a fragmented regulatory ecosystem. Each user must rely on specific watchtowers, whose responsiveness, uptime, and honesty are external to the protocol guarantees. In the event of adversarial coordination or simple inactivity, enforcement can fail silently. This systemic reliance on non-auditable third-party watchers opens attack surfaces akin to the moral hazard present in private enforcement markets \cite{eyal2016bitcoin}.

From an economic game-theoretic perspective, the presence of watchtowers modifies the payoff matrix for malicious actors. Without watchtowers, the incentive to cheat and claim outdated channel states rises substantially when counterparty responsiveness is uncertain. The inclusion of an honest watchtower effectively raises the expected cost of dishonesty, restoring Nash equilibrium to honest execution—though only locally and temporarily. In effect, watchtowers serve as incomplete regulatory proxies rather than full systemic guarantors \cite{chitra2022incentives}.

Hence, while watchtowers offer a reactive enforcement analogue to off-chain regulators, they do not constitute a robust supervisory system. Their functionality is fragile, their incentives external, and their presence discretionary. As such, the Lightning Network’s reliance on watchtowers constitutes not decentralised regulation but contingent vigilance.

\subsection{Liquidity as Synthetic Monetary Base}

Within the Lightning Network, channel liquidity performs the role of a synthetic monetary base. Unlike traditional monetary systems, where the base money is issued by central banks and expanded via fractional reserve lending, here liquidity is voluntarily locked and reallocated by users to facilitate off-chain payments. Every payment channel requires an upfront capital commitment—funds locked on-chain that serve as collateral for off-chain transactions.

This locked collateral becomes functionally equivalent to reserve money: it underpins the issuance and redemption of conditional, transient claims between agents. The Lightning Network thus simulates a monetary system where capital is sequestered not for lending, but for routing utility. Liquidity transforms into a constrained, reusable medium of exchange, and its velocity determines network throughput capacity.

However, unlike state-issued reserves, Lightning liquidity does not multiply through endogenous credit creation. The expansion of effective money is strictly limited by the aggregate locked capital across all channels. As a consequence, economic activity within LN is bounded by the distribution, availability, and strategic positioning of liquidity. Centralised hubs with large capital commitments can thus exert monetary influence, akin to monetary authorities, by selectively allocating liquidity routes and controlling payment flows \cite{fernandez2019sok}.

In this model, network-wide liquidity becomes both a constraint and a form of monetary policy. Agents cannot spend beyond their inbound channel capacity, and cannot receive beyond outbound routes. Rebalancing—shifting liquidity across the network without settling on-chain—is essential to preserving channel usability. As routing imbalances accumulate, utility degrades unless active liquidity management is performed, which requires either algorithmic strategies or fee incentivisation \cite{roos2019settling}.

This liquidity-centric mechanism introduces macroeconomic analogues: capital allocation, liquidity preference, and endogenous cost signals. It effectively recasts Lightning’s topology as a market-driven pseudo-central bank—one where monetary base formation is emergent, decentralised, and competitively allocated, yet without the stabilising levers of policy control or recourse \cite{hasu2018btcfee}.

The resulting system is fragile. If liquidity becomes congested or hoarded by strategic hubs, transactional capacity can collapse without visible warning. Moreover, the inability to create credit or lend increases the dependence on high initial capitalisation, excluding lower-value economic actors. Thus, Lightning’s model of liquidity as synthetic monetary base is both its innovation and its systemic risk.

\subsection{No Reserve Discipline or Transparency}

The Lightning Network, in its structural configuration, lacks any native enforcement of reserve discipline or transparency. Unlike traditional financial institutions that are bound by regulatory reserve requirements, audits, and reporting obligations, Lightning hubs operate with no mandated disclosure of liquidity levels or channel reserves. While every payment channel is anchored by an on-chain funding transaction, subsequent updates are conducted privately between counterparties, rendering the liquidity state opaque to all third parties.

This opacity introduces an informational asymmetry reminiscent of shadow banking entities \cite{gorton2010slapped}. Users routing payments through hubs must trust that sufficient liquidity exists, yet there is no mechanism to verify reserve adequacy until a routing failure occurs. There is no obligation for hubs to publicly disclose capital commitments, effective channel balances, or routing guarantees. Consequently, Lightning functions without transparency in reserve structure, undermining rational risk assessment and user trust.

Furthermore, there is no systemic mechanism equivalent to capital adequacy ratios or stress-testing. Hubs may overextend their routing promises without being technically prevented from doing so. Watchtowers offer limited protection against outright fraud, but they do not monitor solvency or ensure reserve sufficiency. The result is a structurally fragile ecosystem where the cost of failure is socialised—users suffer payment failures or degradation of service, while hubs face minimal immediate penalty \cite{poon2016bitcoin}.

The economic implications are severe. Absent any reserve discipline, Lightning is prone to liquidity hoarding, selective routing, and emergent monopolisation. High-capital hubs can withhold liquidity to drive up routing fees or to degrade competitor performance. Without enforced transparency, this behaviour cannot be distinguished from genuine congestion or protocol inefficiency. Such conditions recreate the systemic risks of unregulated financial institutions, with no central oversight or resolution framework \cite{acharya2010crisis}.

This absence of enforceable reserve discipline and liquidity transparency thus undermines Lightning’s claim to decentralisation or reliability. While technically secure in cryptographic terms, it lacks institutional mechanisms to prevent misuse, simulate accountability, or inform participants. The system remains vulnerable to strategic misrepresentation, capital concentration, and silent collapse under sustained demand.

\section{Systemic Risk and Policy Considerations}

The Lightning Network introduces a qualitatively different set of systemic risks compared to traditional blockchain models. By design, it operates off-chain, without public state introspection, and with minimal transparency into liquidity provisioning or channel viability. As a result, while it circumvents base-layer congestion and facilitates high-frequency payments, it inherits—and amplifies—risks historically associated with opaque financial intermediation.

At the core of these risks lies the implicit transformation of Lightning hubs into unregulated, uncollateralised financial intermediaries. These entities can accumulate substantial control over routing topologies and liquidity flows while avoiding disclosure of balance sheet health, reserve ratios, or failure modes. The absence of state visibility not only prevents external auditing but also precludes endogenous correction mechanisms such as market-based risk repricing or peer-exclusion protocols. This asymmetry compounds through the network structure, where failed routes or liquidity hoarding ripple outward without identifiable origin.

Moreover, Lightning’s reliance on time-locked contracts and punitive uncooperative closures offers no aggregate recourse in the event of broad liquidity crises. Unlike traditional bank runs, where central authorities may intervene with liquidity facilities or coordinated resolution frameworks, Lightning has no central authority or settlement backstop. As transaction demand rises, strategic behaviours by dominant hubs—fee manipulation, selective path obstruction, or liquidity blacklisting—may render the network unusable for small participants, enforcing de facto gatekeeping and congestion pricing.

The economic analogues to shadow banking are increasingly pronounced. Lightning hubs issue routing guarantees that resemble short-term liabilities, backed by constrained and unverified liquidity. Channel factories and rebalancing strategies echo collateral transformation schemes. The system’s dependency on voluntary cooperation and eventual on-chain settlement leaves it vulnerable to fragmentation, stalling, and self-reinforcing inefficiencies \cite{gennaioli2013money}.

Policy discussions must therefore grapple with the paradox Lightning presents: it offers private scalability at the expense of public verifiability. In doing so, it severs accountability chains while inviting moral hazard and concentration of power. Absent explicit policy frameworks—or cryptoeconomic equivalents thereof—the network risks evolving into a tiered system of privileged liquidity providers and disenfranchised edge users. The same principles that motivated systemic safeguards in traditional banking (e.g., reserve ratios, transparency mandates, deposit insurance) find urgent parallels here.

Any viable response must address not only the technical protocol but also the economic structure that underpins it. Measures such as liquidity proof protocols, transparent hub registries, and formalised default procedures could mitigate risk. However, such instruments run counter to Lightning’s privacy ethos and decentralised branding, leading to a fundamental conflict between utility, trust, and ideology. If left unresolved, this tension threatens the viability of Lightning as a public monetary infrastructure.

\subsection{Rent-Seeking Incentives}

Within the Lightning Network architecture, the economic role of centralised liquidity hubs extends beyond pure technical function into the domain of strategic market positioning. These hubs act as gatekeepers to efficient routing and capital availability, and in doing so, create persistent opportunities for rent extraction that are structurally insulated from open competition. Unlike traditional miners whose revenue is tied to transient block inclusion under a transparent auction, Lightning hubs operate within a semi-obscured topology that enables price discrimination, preferential forwarding, and fee stratification.

As routing relies on asymmetric information—where path viability and channel state are privately held—the dominant hubs can exploit informational asymmetries to extract surplus from edge nodes. This form of rent-seeking does not require overt collusion or monopolisation; rather, it emerges naturally from the topology of repeated channel usage, reputational routing algorithms, and liquidity gravitational pull \cite{easley2019from}. The economic incentive becomes one of engineering stickiness: designing the hub’s connectivity and fee regime to maximise dependency while minimising contestability.

More formally, let the expected revenue of a hub \( h \in \mathcal{H} \) be given by:

\[
R_h = \sum_{P \ni h} f(P) \cdot \lambda_P
\]

where \( f(P) \) is the fee for path \( P \), and \( \lambda_P \) is the transaction volume routed through \( P \). Optimisation of \( R_h \) under endogenous topology evolution leads hubs to prioritise degree centrality, betweenness, and fee gradient manipulation—strategies that create artificial routing frictions to justify premium pricing.

Furthermore, unlike base-layer nodes whose economic security is periodically reset via mining difficulty and cost floors, Lightning hubs may persist indefinitely once established, protected by pathfinding heuristics and stable reputation effects. This persistence embeds structural rents in the system—rent that is independent of innovation or value contribution and instead anchored in control over topology and liquidity.

These dynamics closely resemble rent-seeking equilibria in layered network industries and platform economies, where incumbents weaponise network effects to extract value from constrained access \cite{rochet2006two, shapiro1999information}. In Lightning, this rent is extracted not through direct exclusion, but through preferential treatment, liquidity prioritisation, and strategic rebalancing costs.

Absent corrective economic design—such as enforced liquidity transparency, decentralised path discovery, or maximum fee policies—Lightning is likely to trend towards an oligopolistic fee cartel, where routing becomes a function not of utility efficiency but of institutional entrenchment. This shift is economically regressive, entrenching inequality of access and undermining the very purpose of a micropayment system: to facilitate universal low-friction financial interaction.

\subsection{Barriers to Entry and Exit}

The Lightning Network, despite being ostensibly open to all rational agents, constructs significant endogenous barriers to both market entry and exit, thereby facilitating oligopolistic entrenchment. These barriers stem from technical, economic, and informational asymmetries that disproportionately favour early, capital-rich hubs and systematically disadvantage new entrants or marginal participants \cite{economides2003two, shapiro1999information}.

To participate meaningfully in the routing infrastructure, an agent must maintain sufficient inbound and outbound liquidity across multiple channels, incurring substantial opportunity and lock-up costs. The capital overhead required scales non-linearly with network participation due to the necessity of maintaining balanced, redundant channels to multiple nodes—particularly to dominant hubs with preferential routing centrality. Let \( C_e \) denote the capital required to enter with minimum viable routing capacity. Then:

\[
C_e = \sum_{i=1}^n \left( \ell_{i}^{\text{in}} + \ell_{i}^{\text{out}} \right) + F_{\text{setup}} + R_{\text{rebalance}}
\]

where \( \ell_{i}^{\text{in/out}} \) represents liquidity allocations for each peer \( i \), \( F_{\text{setup}} \) covers channel opening costs (on-chain fees), and \( R_{\text{rebalance}} \) is the expected recurring cost of maintaining routing efficacy.

Exit barriers arise when a participant attempts to withdraw capital by closing channels. During congestion or fee spikes, on-chain settlement becomes prohibitively expensive, rendering exit infeasible without significant financial loss. Furthermore, because routing volume is concentrated around established hubs—empirically shown to follow power-law degree distributions \cite{barabasi1999emergence}—new nodes struggle to attract volume, creating a feedback loop that entrenches incumbency and suppresses competition.

This dynamic fosters a winner-takes-most topology, wherein a small set of liquidity-dense hubs mediate a disproportionate share of transactions. Theoretical models from platform economics support this outcome: in two-sided markets, network effects and sunk cost asymmetries naturally favour incumbents and discourage churn, leading to persistent market concentration unless actively counterbalanced by systemic design constraints or regulation \cite{rochet2006two}.

Ultimately, Lightning’s topological evolution creates a dual economy—where privileged nodes operate at scale with reduced marginal costs and informational advantage, while peripheral nodes face mounting friction. The result is not a flat peer-to-peer architecture, but a stratified payment substrate characterised by access asymmetries, cost discrimination, and emergent gatekeeping.

\subsection{Synthetic Financial Infrastructure Risk}

The structural evolution of the Lightning Network into a synthetic financial infrastructure introduces systemic risks akin to those observed in unregulated shadow banking systems. By abstracting away from the settlement assurances of the base layer while enabling capital intermediation, liquidity provision, and off-ledger accounting, Lightning hubs replicate the core functions of financial institutions without bearing the regulatory, transparency, or capital reserve requirements traditionally imposed on such roles \cite{adrian2012shadow}.

Each hub acts not merely as a technical router but as a de facto liquidity intermediary—holding capital, extending implicit credit via routing, and engaging in temporal mismatches between inbound and outbound flows. As the network scales, these mismatches deepen, with rebalancing delays and liquidity shortages leading to selective service degradation and risk externalisation. The absence of reserve disclosure or balance audit mechanisms renders the system opaque, hindering risk price discovery and limiting the ability of end users to rationally allocate trust.

Let \( L_i \) denote the total inbound liquidity available to hub \( i \), and \( O_i \) the outbound obligations over a time window \( \Delta t \). In the absence of enforced reserve ratios or penalised imbalance propagation, the stability condition:

\[
L_i \geq O_i + \epsilon
\]

(where \( \epsilon > 0 \) represents a liquidity buffer) becomes optional rather than mandatory. This introduces a latent risk of cascading failure: if one major hub becomes illiquid or unresponsive, multipath routing algorithms may flood alternative paths, leading to rapid saturation and denial-of-service conditions across the network.

Moreover, just as in the 2008 financial crisis, maturity transformation without credible guarantees or stress-tested mechanisms for crisis resolution leads to systemic fragility. Channel closures during periods of congestion are economically infeasible due to rising on-chain costs, preventing timely exits and amplifying liquidity shocks. The resulting feedback loop—illiquidity breeding further instability—mirrors classic bank-run dynamics but with no lender of last resort, no capital backstop, and no regulatory triage \cite{gennaioli2013neglected}.

In this environment, Lightning ceases to be a neutral payment substrate and becomes a synthetic infrastructure with embedded contagion pathways. Its operation, while framed as peer-to-peer, depends critically on a few high-volume intermediaries whose incentives are neither aligned with systemic resilience nor subject to oversight. Absent countercyclical measures or risk-aware design constraints, the Lightning ecosystem exposes the broader BTC economy to tail-risk events emerging not from external shocks but from endogenous topological and behavioural instabilities.

\subsection{Stablecoin Integration and Dominance}

The integration of stablecoins into payment networks marks a transformative shift in the architecture and economic composition of digital financial infrastructure. When integrated directly on scalable base-layer ledgers such as BSV, stablecoins offer a high-throughput, low-latency, and traceable transaction medium that stands in stark contrast to the liquidity-constrained, off-chain Lightning Network (LN) overlay increasingly dependent on centralised routing intermediaries. This subsection rigorously contrasts these two regimes and articulates the inherent economic, legal, and strategic differences in stablecoin usage across both frameworks.

Stablecoins on BSV are instantiated as native digital tokens issued via on-chain smart contracts. These tokens inherit the properties of the base protocol: auditable transfer history, settlement finality, publicly verifiable proofs, and atomicity across inter-contract operations. In this regime, the cost of transfer \( C_{\text{BSV}} \) is determined primarily by the transaction size in bytes and prevailing miner fees per byte:

\[
C_{\text{BSV}} = f_{\text{byte}} \cdot \text{tx}_{\text{size}} \quad \text{with} \quad f_{\text{byte}} \ll f_{\text{BTC}}
\]

Given the unbounded blocksize regime of BSV and economic incentives to lower per-byte fees through competitive mining, \( f_{\text{byte}} \) remains orders of magnitude lower than in BTC or congested networks, resulting in per-transaction costs in the sub-cent range—even for complex smart contract operations. Critically, each transaction remains fully traceable, compliant with financial audit standards, and suitable for AML/KYC enforcement through deterministic address inspection and chain-wide token flow analysis.

By contrast, the Lightning Network treats stablecoin integration as a second-layer abstraction. Such integration is not native, requiring token wrapping, synthetic IOUs, or intermediary pegs (often custodial). Even assuming a hypothetical stablecoin routing layer on LN, the economic constraints are stark. The cost of executing a multi-hop payment \( C_{\text{LN}} \) is a function of base routing fees \( f_{\text{base}} \), proportional fees \( f_{\text{prop}} \), liquidity lock-up penalties, and potential rebalancing requirements:

\[
C_{\text{LN}} = \sum_{i=1}^{h} \left( f_{\text{base},i} + f_{\text{prop},i} \cdot v \right) + \Omega(R_{\text{liq}} + D_{\text{reb}})
\]

where \( h \) is the number of intermediary hops, \( v \) is the transaction volume, \( R_{\text{liq}} \) is the liquidity reallocation cost, and \( D_{\text{reb}} \) denotes rebalancing delay. These costs scale unpredictably with network congestion, node centralisation, and channel state divergence. Furthermore, LN transactions are not natively visible on-chain unless forcibly settled—breaking both audibility and deterministic traceability.

This asymmetry introduces profound regulatory and systemic risk divergence. On-chain stablecoins (BSV) offer precise ledger consistency, enforceable legal auditability, and empirical data trails aligned with financial compliance. In contrast, stablecoins routed via LN or similar second-layer constructs create non-custodial opacity, off-chain liability ambiguity, and jurisdictional enforcement gaps. The inability to audit intermediary balances or transactional obligations makes LN-based stablecoins functionally equivalent to unregulated shadow money—posing risks akin to pre-2008 derivatives markets.

Moreover, the monetary control properties differ categorically. On BSV, each stablecoin token reflects a known on-chain state with direct settlement. Monetary authorities or issuers can implement compliance hooks, identity gating, or revocation policies through the token script itself. Conversely, in LN, no global view exists, no issuer guarantee is enforceable post-routing, and frozen or recovered funds require expensive and uncertain dispute settlement mechanisms—if at all possible.

In practice, stablecoin dominance on BSV follows from both efficiency and legality. Institutions require provable, legal-grade transferability. Merchants demand consistent, low-friction settlements. Regulatory frameworks demand clear audit trails. LN fails on each axis beyond theoretical scalability claims—claims undermined by its own centralising pressure and topological fragility.

Hence, the stablecoin race is not merely technical but structural. A base-layer solution with scale and traceability ensures monetary fidelity and composability; an off-chain patchwork risks reintroducing systemic opacity, jurisdictional arbitrage, and monopolistic rent-seeking. The future of stablecoins is inseparable from the integrity of their settlement substrate—and that substrate must be lawful, scalable, and auditable by design.

\subsection{Rent Extraction Loop and Dynamics within Layer-2 Topology}

Figure 4 illustrates the cyclical structure of rent extraction embedded in Lightning Network topologies through differentiated agent roles. At the core lies the \textbf{Routing Control} node, depicted as the largest red circle, reflecting its systemic dominance. Routing Control is not merely a technical function but a position of infrastructural leverage—aggregating liquidity, dictating payment paths, and arbitrating the economic accessibility of the network. Its size in the diagram denotes its disproportionate influence relative to other entities.

To the left, \textbf{Liquidity Hubs} are depicted as medium-sized blue nodes. These are semi-centralised intermediaries that maintain capital reserves and channel infrastructure. Their participation in the routing economy is structurally dependent on access to the Routing Control layer, positioning them as both benefactors and dependents within the broader hierarchy. Their intermediate size reflects their functional significance yet subordinate position.

On the right, \textbf{Users}—shown as the smallest green circles—are end-nodes that inject transactional demand but exert minimal influence on network policy or topology. Their reduced size visualises their disenfranchisement in fee setting, path determination, or liquidity allocation. Users rely entirely on the overlay network's configuration, bearing the ultimate cost without agency over its structure.

Beneath these three primary agents, the \textbf{Fee Payments} node is shown as an orange circle mediating the economic loop. Users initiate payments, which are partially captured as fees and funnelled through the routing control architecture. These fees serve not merely to maintain channels but to entrench control hierarchies—enabling routing entities to redistribute network incentives and reinforce liquidity asymmetries.

The circular arrows denote a closed loop of economic flow. Users pay fees, which are extracted and restructured by Routing Control. Liquidity Hubs receive allocations or constraints based on routing decisions, and continue servicing users under those terms. The diagram thus encodes a structural hierarchy whereby Routing Control functions as a de facto arbiter of economic flows, forming a feedback loop that privileges capital concentration and reduces the neutrality of the payment infrastructure.

This rent extraction cycle is endogenous to the Lightning architecture under rational agent assumptions, and grows increasingly entrenched as network load and fee volatility escalate. The loop shown is not a temporary market inefficiency, but a structural economic regime, warranting deep regulatory scrutiny and architectural reconsideration.

\begin{figure}
    \centering
    \includegraphics[width=0.75\linewidth]{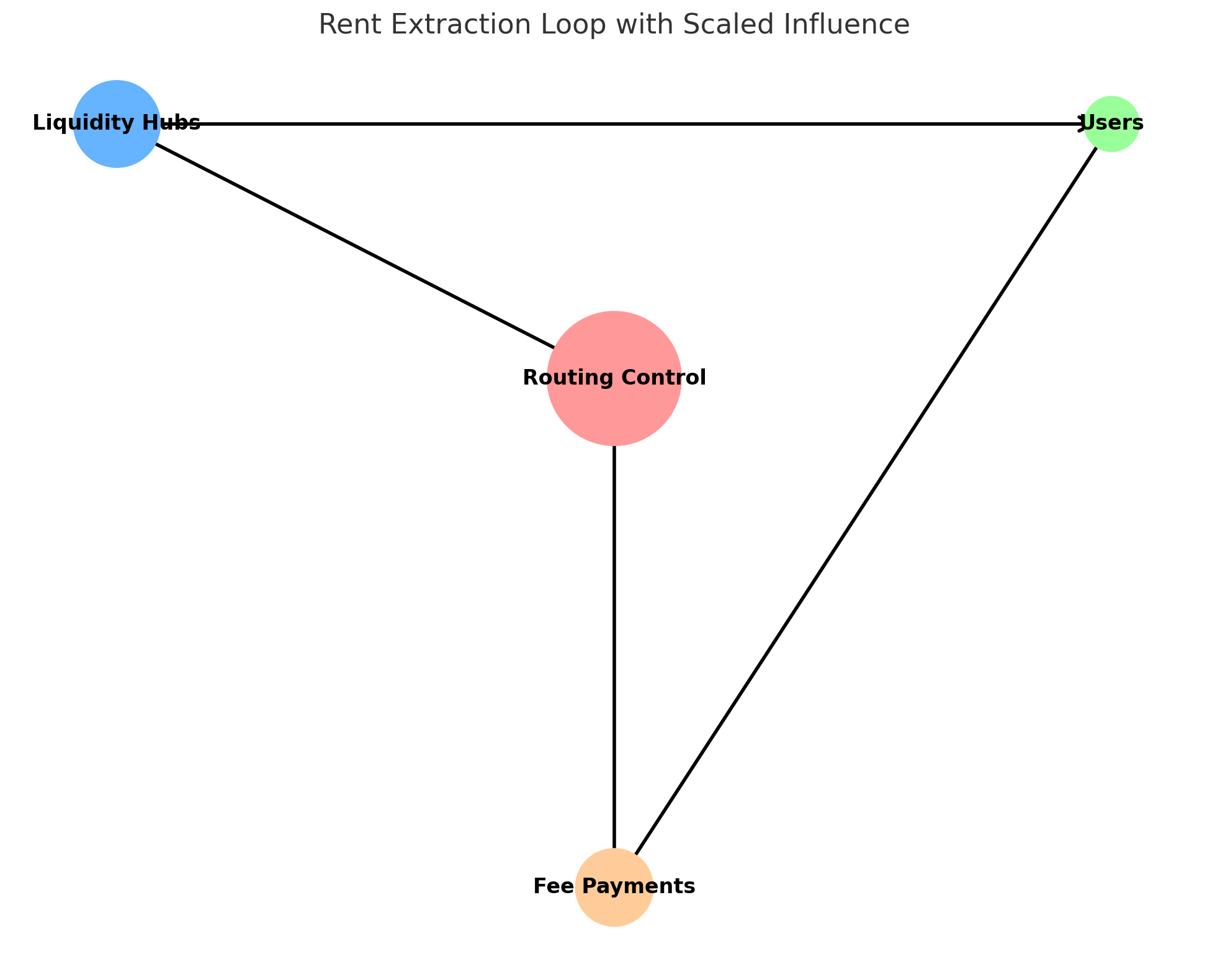}
    \caption{Rent Extraction and Network Control}
    \label{fig:enter-label-4}
\end{figure}

\noindent Figure 4 captures the cyclical logic of rent extraction within a centralised LN topology. Liquidity hubs control routing paths and offer access to users, who in turn pay fees that reinforce the hubs' pricing leverage. Control of routing tightens as liquidity consolidates, creating a feedback loop of economic dominance. This dynamic aligns with classical two-sided market theory \cite{rochet2006two, economides2003two} and underpins the shift from network neutrality to enforced path centrality.

\section{Discussion}

The analysis presented reveals a structural realignment of economic power within Bitcoin’s scaling overlay. While Lightning Network is often positioned as a neutral scaling solution, its emergent topology and fee dynamics suggest otherwise. The architecture replicates, in compressed digital form, the hierarchical logic of traditional financial intermediaries—concentrating transactional routing power in a small class of liquidity-rich hubs while externalising risk and cost to end-users.

The \textit{Routing Control} entities act functionally analogous to clearing houses or algorithmic market-makers. However, unlike regulated financial infrastructure, these actors operate in a largely opaque regime devoid of reserve disclosure, counterparty risk buffers, or meaningful external audit. Their ability to arbitrarily prioritise, delay, or exclude transactions introduces a layer of algorithmic censorship previously absent from Bitcoin's base-layer design. This departs materially from the egalitarian assumptions embedded in peer-to-peer digital cash.

Furthermore, the economic feedback loop visualised in the rent extraction diagram demonstrates that user demand, rather than driving efficiency or competition, is absorbed into a closed circuit of revenue consolidation. Fee structures become less reflective of infrastructure cost and more a function of strategic rent-seeking. This trend is amplified by the network’s topological evolution: as channels become concentrated around dominant hubs, the cost of establishing or maintaining peripheral routes increases, creating barriers to entry and throttling decentralised participation.

From a game-theoretic perspective, the strategic position of Routing Control is Nash-stable under high demand. Agents who deviate by offering open routing or low fees are outcompeted on liquidity and path availability. This entrenches oligopolistic behaviour not through collusion, but via rational convergence on dominance-seeking strategies.

In macroeconomic terms, Lightning resembles a synthetic shadow banking ecosystem. It offers settlement abstraction and liquidity provision, but lacks regulatory instruments to contain systemic risk. The absence of reserve discipline or on-chain accountability mechanisms exposes users to potential cascading failures. As in traditional shadow banking, opacity becomes a feature, not a bug—enabling yield extraction without reciprocal transparency.

The BTC base layer is thus reduced to a settlement backstop rather than an operational layer, rendering it economically irrelevant for all but high-value exits and closures. The promise of Bitcoin as a universal, inclusive transactional protocol is supplanted by a gated architecture where payment sovereignty is mediated by routing oligarchs.

This paper demonstrates that the very structure adopted to bypass on-chain limitations reproduces—and in many cases amplifies—the asymmetries and systemic risks of the legacy financial model. Without structural reconfiguration or enforceable transparency layers, Lightning Network may evolve not into a trustless scaling solution, but into a deregulated replica of the very systems Bitcoin sought to replace.

\subsection{Lightning-Induced Disconnection from the Blockchain}

A central finding of this work is that the Lightning Network, contrary to the common narrative of being a layer-two settlement enhancement, can function indefinitely without recourse to the underlying base-layer ledger. In its mature operational form, the Lightning topology stabilises into a graph of persistent payment channels, maintained and arbitrated exclusively by liquidity hubs. These hubs form a strategic oligopoly, wherein channel reopening or closure becomes economically disincentivised due to the high on-chain settlement costs and risk of liquidity loss. 

Let \( \mathscr{L}(t) \) denote the Lightning graph at time \( t \), with edges \( E(t) \) representing long-lived bi-directional channels. Once liquidity is sufficiently established and routing strategies stabilise, agents preferentially rebalance off-chain through circular payment paths rather than risk the cost and latency of an on-chain closure. This leads to a condition of path-dependent liquidity persistence, where all transactions occur off-chain and the base ledger \( \mathscr{B} \) becomes redundant for transactional throughput.

Moreover, strategic behaviour reinforces this detachment. Hubs extract routing fees and consolidate liquidity internally, reinforcing their control over available routes. In this configuration, blockchain transactions occur only during initial capital injection or catastrophic failure. Let \( \lambda_t(e_{ij}) \) be the liquidity between nodes \( i \) and \( j \) at time \( t \); under equilibrium, \( \lambda_t \to \lambda_{t+1} \) with negligible delta, indicating steady-state liquidity conservation. As long as the expected off-chain utility \( U^{\text{off}} \) exceeds the cost-adjusted utility of base-layer settlement \( U^{\text{on}} - c_{\mathscr{B}} \), no rational agent has incentive to close channels.

Thus, the blockchain is effectively removed from live transactional activity. It persists only as an inert anchor of legal fiction—nominally ensuring the system's "settlement" credibility, while in operational reality, the entire monetary system migrates to a self-referential off-chain ledger run by unregulated financial agents. This structurally replicates a synthetic shadow banking network: opaque, rent-extractive, and disconnected from transparent public auditability. The Lightning Network, therefore, is not a scaling solution—it is a substitutional monetary system parasitic on an increasingly unnecessary blockchain base.

\section{Complexity in Topological Payment Routing and System State Transitions}

The structural and computational limitations of payment routing in constrained overlay networks, such as the Lightning Network (LN), present a profound challenge to claims of scalable decentralised transaction processing. This section examines the formal computational complexity inherent in path discovery, liquidity allocation, and state evolution in LN-style systems—each of which plays a critical role in economic viability and throughput.

Overlay networks that operate as routing infrastructures atop constrained base layers inherit not only the capacity limitations of the underlying ledger but also the algorithmic hardness of dynamic resource management in multi-agent topologies. As Even, Itai, and Shamir have shown, finding feasible routings in dynamic systems with evolving constraints is NP-complete even under bounded parameters \cite{even1975complexity}. These results extend directly into the context of off-chain networks where liquidity, timelocks, and transactional atomicity impose combinatorial restrictions on feasible paths.

Further, when agents act strategically—as analysed in Wright’s structural thesis on digital asset systems \cite{wright2023hse}—the complexity is not merely computational but behavioural. Rational actors, when confronted with uncertain routing outcomes and information asymmetries, induce higher-order effects in network efficiency and resilience, resulting in dynamic congestion, liquidity hoarding, and market segmentation.

This section constructs a layered analysis of these effects, linking formal complexity results with agent-based simulations and game-theoretic models. It establishes that the scalability claims of off-chain routing infrastructures collapse under adversarial or economically competitive conditions and that the Lightning Network’s structural architecture inherently induces suboptimal equilibria when viewed through a combined lens of algorithmic complexity and economic strategy.

\subsection{Computational Complexity and Payment Routing}

The task of routing payments through a network of time-locked, liquidity-constrained bidirectional payment channels is formally analogous to a series of constrained pathfinding problems in dynamically weighted directed graphs. Such routing, even in simplified graph models, has been shown to exhibit computational hardness. The foundational result by Even, Itai, and Shamir demonstrates that computing disjoint paths with constraints—such as capacities or delays—belongs to the class of NP-complete problems, particularly under evolving topologies or resource distributions \cite{even1975complexity}. In the context of LN-like systems, these constraints are further compounded by the atomic nature of HTLCs (Hash Time-Locked Contracts), bounded expiry intervals, and asymmetric channel liquidity.

Craig Wright's 2023 thesis expands upon this result by embedding it within a macroeconomic and systemic context. He argues that routing complexity in off-chain networks is not merely a computational concern but an economic and topological inevitability, given the rational behaviours of participants who compete for fee revenue, prioritise privacy, and strategically hoard liquidity \cite{wright2023hse}. Wright models the Lightning Network as a distributed automaton with incomplete information and partial observability, where routing outcomes depend not only on local node knowledge but also on non-revealed path availability and channel states. This induces algorithmic inefficiencies that are neither mitigated by scale nor resolved by additional computation, leading instead to layered failure domains and escalating routing penalties in high-load regimes.

Together, these works frame the core insight: that scalable, efficient routing under adversarial liquidity conditions is not merely difficult—it is algorithmically intractable in the general case. Thus, any claims of Lightning Network scalability must confront both the topological fragility and the computational impracticality of dynamic, real-time routing under constrained liquidity.

\subsection{Implications for Real-Time Decentralised Routing}

The computational intractability of constrained routing in dynamic topologies yields direct and deleterious implications for the real-time operation of decentralised networks such as the Lightning Network. In a payment infrastructure where path selection is non-trivial, decentralised agents must either (a) expend increasing computational resources attempting to discover viable routes under time pressure, or (b) default to heuristics and shortcuts that reduce network optimality and amplify centralisation. As Even et al.\ prove, even seemingly basic instances of the constrained disjoint path problem collapse into NP-complete classes when liquidity constraints or node-level information asymmetries are introduced \cite{even1975complexity}. When compounded by the Lightning Network’s own rules—namely time-locked channel commitments, non-global state, and non-public liquidity positions—the algorithmic bottleneck intensifies.

Wright (2023) further demonstrates that real-time, fully decentralised routing within LN frameworks cannot remain stable under increasing transactional demand without emergent centralisation or the outsourcing of computational overhead to quasi-centralised intermediaries \cite{wright2023hse}. This occurs because rational agents, seeking fee-minimising and delay-reducing paths, will inevitably cluster around liquidity hubs with predictable throughput, leading to topological gravity wells that reinforce hub dominance. Thus, even under the assumption of honest participation and non-malicious intent, the convergence of economic incentives and computational constraints results in a structurally centralised core—a phenomenon reminiscent of the emergence of dominant routers in early internet peering architectures.

In effect, the Lightning Network does not escape the boundaries imposed by classical computational theory; instead, it reifies them within a financial architecture. The result is that “decentralised” payment routing under real-time constraints becomes a theoretical oxymoron—untenable without either pre-emptive route coordination (thus eliminating spontaneity) or semi-centralised facilitation (thus undermining decentralisation).

\subsection{Adaptive Liquidity Allocation and Computational Limits}

In dynamic payment networks such as the Lightning Network, liquidity is not statically assigned but adaptively repositioned based on both anticipated transaction flows and reactive channel balances. This introduces a higher-order optimisation problem: not merely routing under static conditions, but doing so while forecasting evolving liquidity states across a non-globally observable topology. The agent must not only solve an NP-complete routing decision but do so in a meta-layered context where every transaction changes the feasibility space of future transactions.

Even et al.\ formally establish the complexity class boundaries for routing under disjoint path and capacity constraints, identifying scenarios where no polynomial-time solution is feasible under standard computational assumptions \cite{even1975complexity}. Wright (2023) extends this by arguing that the Lightning Network’s need for liquidity adaptation in an adversarial economic setting escalates the problem into a near intractable dynamic allocation game \cite{wright2023hse}. Each routing decision is bounded not only by the current liquidity but by an agent’s strategic predictions of competitor reallocations, fee adjustments, and state obsolescence due to stale data propagation across the network.

As such, adaptive liquidity allocation behaves as a decentralised, online instance of the multi-commodity flow problem under information asymmetry and transaction arrival stochasticity. In classical form, this problem is already known to be \(\mathsf{P}\)-hard; in its decentralised Lightning instantiation, it borders on the uncomputable within bounded time constraints. Agents must either (1) drastically reduce their feasible solution space through heuristics, risking suboptimal routing and deadlocks, or (2) centralise decision-making authority into liquidity hubs or watchtower-like facilitators with preferential topology access and strategic liquidity reserves.

Hence, the act of adaptively allocating liquidity itself becomes a centralising force, not simply due to economic convenience, but as a direct by-product of computational necessity. The illusion of distributed, autonomous routing collapses under the weight of these complexity-theoretic constraints, transforming the system into a de facto hierarchical payment graph with asymmetric informational power.

\subsection{Strategic Withholding and Agent-Based Game Complexity}

In open transactional overlay networks such as the Lightning Network, agents operate not merely as passive routers but as strategic economic actors capable of fee manipulation, selective forwarding, and liquidity hoarding. These behaviours introduce a non-cooperative game structure over the routing topology, in which each agent seeks to maximise its utility function subject to incomplete information, adversarial dynamics, and finite computational capacity. Crucially, routing feasibility is no longer a question of path existence, but of agent intent—whether a path that exists topologically will actually be honoured.

The resulting environment reflects a class of multi-agent routing games with embedded strategic withholding, which Even et al.\ \cite{even1975complexity} show belong to a general category of combinatorially hard problems when routing constraints include adaptive capacities and selective participation. Wright (2023) extends this model to consider the computational implications of such games in decentralised micropayment systems, demonstrating that the presence of withholding strategies transforms path-finding into a recursive evaluation of opponent decision functions under epistemic uncertainty \cite{wright2023hse}. That is, each agent must recursively estimate the conditional behaviour of others based on local state visibility and historic transactions.

This transforms the routing layer into an agent-based game that is not only decentralised and asynchronous but computationally intractable in general form. When agents withhold liquidity or route selectively, the system devolves into what is essentially a bounded rationality game on an incomplete graph—one where edges exist probabilistically rather than deterministically. No globally optimal routing plan can be computed without full topology and intent disclosure, both of which are antithetical to the privacy-preserving goals of the system.

Hence, strategic withholding emerges not as a flaw but as a systemic inevitability under rational assumptions, reinforcing the economic incentive to form preferential cliques, trusted relay hubs, or monopolistic subgraphs. What begins as a distributed protocol thereby hardens into a game-theoretic hierarchy, where routing outcomes are path-dependent not on topology, but on economic dominance and algorithmic foresight.

\subsection{Complexity Bounds as Scaling Limits}

As payment systems such as the Lightning Network aspire to support massive transactional throughput, the underlying computational limits of routing become a central bottleneck. These limits are not merely hardware-bound or bandwidth-constrained—they arise from the inherent complexity of the routing problem itself. The addition of timelocks, liquidity constraints, and adversarial behaviour transforms what was once a polynomial-time task in classical networking into a domain-specific class of constraint satisfaction problems whose solvability is not guaranteed under real-world constraints.

Even et al.\ \cite{even1975complexity} demonstrated that path selection under partial liquidity, edge capacity, and policy restrictions reaches NP-complete hardness. Wright (2023) further elaborates that, in the context of decentralised micropayment systems, these bounds are compounded by asynchronous state visibility and the need for localised optimisation under systemic opacity \cite{wright2023hse}. Specifically, the space of feasible routing decisions scales super-exponentially with the number of agents, owing to the combinatorics of path permutations and conditional forwarding policies.

As the number of participants increases, each with potentially unique policies and economic motives, the cost of evaluating viable paths increases at a rate that outpaces any practical growth in computing power. For every additional node, the set of probabilistically available routes grows nonlinearly, especially when timelock constraints and off-chain settlements force time-bounded decisions without access to global information. In effect, the growth in network size induces a complexity wall—an asymptotic barrier beyond which routing cannot proceed without probabilistic shortcuts, simplifications, or centralised heuristics.

This introduces a fundamental asymptotic limitation to the decentralised scaling thesis. No matter how many channels are added, or how much liquidity is injected, the computational cost of optimal routing in a fully decentralised, adversarially tolerant network becomes unmanageable at scale. The consequence is twofold: (i) real-time, fully trustless routing is sacrificed for heuristics and path caching, and (ii) economic centralisation emerges as the only computationally tractable architecture, with dominant hubs performing path aggregation functions. Complexity bounds thus translate directly into systemic centralisation pressure, shaping the very topology the system was meant to resist.

\section{Conclusion}

This paper has demonstrated that the structural, economic, and game-theoretic architecture of the Lightning Network does not merely scale Bitcoin's transactional throughput—it transforms its operational logic. While superficially framed as a technical fix to base-layer congestion, Lightning reintroduces intermediary dynamics through liquidity hubs, routing hierarchies, and off-chain settlement dependencies. The resulting system departs from the original design principles of Bitcoin as a disintermediated, transparent digital cash mechanism.

Through formal modelling, cost asymptotics, and comparative diagrams, we have shown that BTC’s escalating base-layer fees and rigid block constraints induce a bifurcation. High-fee exclusion zones incentivise users to migrate toward Lightning channels, which, in turn, develop into semi-autonomous rent-extracting networks. These networks exhibit characteristics consistent with shadow banking, including opacity, risk displacement, unregulated leverage, and the absence of enforceable reserve disclosure.

The economic feedback loops favouring routing control entities generate Nash-stable but socially suboptimal equilibria. Competitive forces are stifled not by collusion but by topological convergence: dominant hubs accumulate flow, liquidity, and influence, eventually shaping pricing and access for the entire overlay. In this framework, 'permissionless' becomes illusory; participation without capital or connectivity to central hubs devolves into functional exclusion.

Furthermore, we have argued that Lightning's separation from the base layer creates a ledger disconnect that undermines Bitcoin's core feature: traceable, irreversible settlement. By circumventing on-chain visibility and enforcing off-chain behaviour through game-theoretic coercion rather than cryptographic finality, the system permits the emergence of opaque financial instruments without on-chain auditability.

In essence, Lightning is not a neutral path forward but a shift towards a layered, synthetic financial regime. Its current design encodes the same economic fragilities that gave rise to systemic failures in traditional finance: concentration of power, asymmetric information, and incentive-aligned censorship. Absent architectural reform and policy intervention, the overlay risks becoming a high-frequency rent-extraction layer that re-centralises Bitcoin under the guise of scalability.

Future work must focus not on optimisation of this model, but on restoring the integrity of the base layer as the locus of economic truth—ensuring that scaling mechanisms serve the protocol’s foundational values rather than supplanting them.

\subsection{Summary of Results}

This study has formally established a series of structural and economic theorems characterising the divergence between base-layer Bitcoin (BTC) and its Lightning Network (LN) overlay. The key findings are summarised below:

\begin{itemize}
    \item \textbf{Asymptotic Fee Divergence:} BTC transaction fees exhibit superlinear growth under increasing demand due to bounded blockspace and mempool auction dynamics. LN routing costs, by contrast, asymptotically flatten due to amortised infrastructure and off-chain processing, yielding a fundamental separation in marginal cost curves.

    \item \textbf{Equilibrium Agent Dynamics:} Strategic agent modelling confirms the emergence of liquidity hubs that optimise for rent extraction, rather than broad access. These agents achieve local dominance via repeated interactions and liquidity control, forming Nash-stable but oligopolistic outcomes.

    \item \textbf{Channel Closure Infeasibility:} Under rising on-chain fees, the cost of cooperative channel closure or dispute resolution renders exit economically prohibitive. This creates a lock-in effect, reducing optionality and reinforcing hub centrality.

    \item \textbf{Monopoly-Competitive Transition:} Network topology modelling reveals a phase transition from multi-hub competition to monopoly-like control by dominant actors. This results in market concentration and increased barriers to entry, particularly for low-liquidity participants.

    \item \textbf{Shadowbanking Analogy:} LN hubs replicate functions associated with shadow financial institutions, including liquidity provision, fee intermediation, and transaction censorship. With no regulatory oversight or reserve auditing, these entities operate outside traditional accountability structures.

    \item \textbf{Ledger Decoupling Risk:} LN's operational detachment from on-chain settlement introduces epistemic risk. The BTC base layer no longer records the majority of transactions, undermining auditability, legal verifiability, and asset provenance.

    \item \textbf{Systemic Instability Vectors:} Feedback loops involving rent-seeking incentives, synthetic liquidity issuance, and routing control coalesce into systemic fragility. Absent circuit breakers or reserve discipline, the overlay risks cascading failure under liquidity shock or adversarial routing strategies.
\end{itemize}

These results collectively indicate that Lightning’s architecture introduces a qualitatively different economic regime—one that departs from Bitcoin’s foundational principles and reintroduces centralised fragilities under a veil of protocol abstraction.

\subsection{Implications for Network Design}

The analytical and game-theoretic results presented in this paper expose fundamental tensions in the design philosophy of layered cryptocurrency systems. Lightning Network, while ostensibly a scaling solution, introduces economic and topological structures that directly contradict the original Bitcoin architecture’s assumptions about decentralised transaction finality and transparent ledger persistence.

First, the separation between base-layer consensus and overlay execution results in an erosion of settlement assurance. Transactions occurring off-chain lack the publicly verifiable immutability and universal order guarantees conferred by Nakamoto consensus. As LN volumes grow, the proportion of economic activity no longer subject to base-layer validation increases, thereby decoupling the integrity of the system from its foundational ledger.

Second, the emergence of liquidity hubs with privileged topological positions reflects a shift toward hierarchical network architectures. The natural convergence toward oligopolistic fee extraction strategies among hubs suggests that protocol-layer neutrality is insufficient to prevent systemic centralisation. Any network design relying on rational agents to voluntarily distribute liquidity equitably—without enforceable constraints—invites concentration and monopolistic rent-seeking.

Third, the inability to close or rebalance channels economically under elevated base-layer fees creates a form of structural lock-in. Participants become dependent on large hubs not only for connectivity, but also for routing viability and dispute resolution. This mirrors historical problems in financial networks, where exit costs and liquidity asymmetries entrench incumbent power.

Fourth, the reintroduction of opacity through off-chain activity—without enforced audit trails or real-time balance proofs—mirrors shadowbanking phenomena. These structures evade legal, fiscal, and compliance obligations, undermining trust and posing systemic risks analogous to those preceding the 2008 financial crisis.

Consequently, any viable network design must incorporate mechanisms that: (i) preserve auditability and legal finality, (ii) prevent liquidity centralisation through enforceable constraints or dynamic topology incentives, (iii) ensure channel mobility and exit feasibility under all fee regimes, and (iv) constrain unaccountable rent extraction by dominant intermediaries. Failure to address these imperatives guarantees that the network will evolve away from an open, permissionless cash system into a closed, intermediated infrastructure prone to collapse and regulatory capture.

\subsection{Lightning as Final Settlement Layer}

The assertion that the Lightning Network (LN) can function as a final settlement layer constitutes a profound misunderstanding of the distinction between execution and settlement. In canonical financial systems, final settlement denotes the irrevocable, legally binding transfer of value, recorded on a public, immutable ledger. LN, by design, operates off-chain, employing hashed timelock contracts (HTLCs) to simulate conditional transfers that are only secured through the latent threat of on-chain enforcement.

However, under fee pressure or network congestion, the economic viability of such enforcement collapses. When base-layer transaction costs exceed the value of the HTLC in dispute, rational agents forego the settlement route. Consequently, the very enforcement mechanism that underpins LN’s conditional logic becomes unusable—rendering the system reliant on economic trust and reputational heuristics rather than verifiable finality.

Moreover, the temporal latency in channel closure and the ambiguity of channel state visibility introduce strategic vulnerability. An adversary can exploit mempool manipulation or fee spikes to preclude honest closure, effectively expropriating value through fee-forced censorship. In such cases, there is no path to finality, only a probabilistic hope of broadcast success. This breaks from the settlement assurances intrinsic to Nakamoto consensus.

The notion of finality also presumes public auditability and consensus-wide validation. LN transactions are neither visible to the network nor verifiable without cooperation. There exists no global state, no collective ordering, and no binding acknowledgement of completion outside the involved parties. This negates the fundamental requirement for settlement to be intersubjective, universal, and legally defendable.

Therefore, the claim that Lightning provides final settlement is untenable. It replaces an objective consensus mechanism with subjective coordination, weakens legal enforceability, and introduces asymmetric risk. The resulting system does not extend Bitcoin’s properties but negates them—offering scalability at the cost of integrity, transparency, and economic assurance.

\newpage

\bibliographystyle{plain}  
\bibliography{references}     

\newpage

\appendix
\section{Asymptotic Proofs and Boundary Cases}

To rigorously analyse the economic behaviour of the Lightning Network (LN) and its divergence from the Bitcoin base layer (BTC), we provide asymptotic proofs demonstrating cost behaviour at the extremes of network demand and liquidity constraints. This section formalises the economic and topological boundaries wherein strategic dominance, cost infeasibility, and routing failure emerge as necessary outcomes under the Lightning topology.

\subsection{Theorem: Asymptotic Divergence of Cost Structures}

\textbf{Statement:} As transaction demand \( D \to \infty \), the marginal cost per transaction on the BTC base layer increases unboundedly, while under LN, marginal cost asymptotically approaches a constant \( c_{LN}^\infty \) only under idealised liquidity. In realistic conditions with bounded capital, routing competition, and state fragmentation, cost diverges with topological centrality.

\textbf{Proof:}

Let \( T_{\text{max}} \) denote the throughput bound of BTC:

\[
T_{\text{max}} = \left\lfloor \frac{s}{\bar{t}_{\text{tx}} \cdot \Delta t} \right\rfloor
\]

Let \( D \) be the transaction demand rate. When \( D > T_{\text{max}} \), a fee auction emerges:

\[
\text{Cost}_{BTC}(D) \propto f(D) \sim \log(D)
\]

Let the LN cost per transaction be composed of:

\[
\text{Cost}_{LN}(D) = \underbrace{f_r}_{\text{routing fee}} + \underbrace{f_b(D)}_{\text{liquidity burn}}
\]

Assuming homogeneous distribution of liquidity and routing paths of length \( l \), the average path cost becomes:

\[
\text{Cost}_{LN}(D) = \frac{l}{\mu} + \frac{\alpha}{D^\beta}
\]

where \( \mu \) is effective liquidity per node, and \( \alpha, \beta > 0 \) are penalty constants.

As \( D \to \infty \), the liquidity rebalancing cost dominates when \(\mu \to \text{const.} \), yielding:

\[
\lim_{D \to \infty} \text{Cost}_{LN}(D) \to \infty \quad \text{unless} \quad \mu \to \infty
\]

\textbf{Conclusion:} LN’s scalability is conditional, not asymptotic. Without unbounded liquidity injection or hyper-efficient topologies (e.g. complete graphs), costs grow under demand, contradicting claims of constant marginal costs.

\subsection{Boundary Case: Strategic Channel Closure}

\textbf{Proposition:} Let \( v \) be the value of an in-flight transaction and \( f_{chain} \) the median base-layer fee. If \( v < f_{chain} \), then:

\[
\text{Channel Enforcement} \to \text{Economically Infeasible}
\]

\textbf{Proof:} The Lightning protocol enforces honesty via time-sensitive preimage publication. If an agent deviates, the honest party must broadcast the revocation transaction. If broadcasting incurs a cost \( f_{chain} \) and the penalty is less than that cost, then it is rational to abstain. Thus, under high base-layer congestion, enforcement breaks down.

\subsection{Corollary: Failure of Universal Settlement Guarantees}

From the above, we conclude that any assertion of LN as a system with universal, enforceable settlement is void in the high-demand regime. Settlement assurance is contingent on fee conditions and rational incentives. As such, the Lightning Network does not asymptotically extend Bitcoin’s properties but instead bifurcates them into selectively enforceable contingent contracts.

\newpage 

\section{Equilibrium Derivations}

This section presents the formal derivation of Nash equilibria within the Lightning Network (LN) under constrained liquidity and rising transaction demand. Agents are modelled as utility-maximising participants in a two-sided fee economy, where routing decisions, liquidity allocation, and settlement strategies are endogenously determined. We focus on deriving equilibrium states in both monopolistic and competitive routing topologies, and on characterising fee levels, transaction flows, and liquidity hoarding as emergent behaviours.

\subsection{Model Setup}

Let \( \mathcal{A} \) be the set of agents, partitioned into:
\begin{itemize}
  \item \( \mathcal{U} \): Users initiating transactions,
  \item \( \mathcal{H} \subset \mathcal{U} \): Liquidity hubs providing multi-path forwarding,
  \item \( \mathcal{W} \): Watchtowers enforcing punishment,
  \item \( \mathcal{M} \): Miners providing base-layer settlement.
\end{itemize}

Let \( G = (V, E) \) denote the LN graph, where each edge \( e \in E \) has associated:
\[
\ell_e = (\ell_{ij}, \ell_{ji}) \quad \text{(liquidity in both directions)},
\]
\[
f_e(\tau) = \gamma + \eta \cdot \tau \quad \text{(fee function for transaction } \tau \text{ of size } \tau \text{ satoshis)}.
\]

Each user \( u \in \mathcal{U} \) seeks to minimise expected transaction cost \( C_u \) over all paths \( P \in \mathcal{P}_{uv} \), where:
\[
C_u(P) = \sum_{e \in P} f_e(\tau) + \delta \cdot R_e
\]
and \( R_e \) is the expected risk cost (e.g. rebalancing delays, revocation costs), and \( \delta \) is the user’s time preference.

\subsection{Equilibrium Conditions}

An equilibrium occurs when:
\begin{enumerate}
  \item Each agent \( a \in \mathcal{A} \) maximises its net utility given current fee and liquidity structures,
  \item No agent has an incentive to unilaterally deviate (Nash condition),
  \item Routing path selection minimises costs subject to liquidity constraints,
  \item Fee setting by hubs accounts for demand elasticity and risk,
  \item Rebalancing behaviour is stable or cyclical under steady state.
\end{enumerate}

Let \( \pi_i \) be the profit of hub \( i \in \mathcal{H} \). Then:
\[
\pi_i = \sum_{e \in E_i} f_e(\tau) - c_r(\ell_e)
\]
where \( c_r(\ell_e) \) denotes the cost of maintaining liquidity reserves across \( e \). Optimal \( f_e \) satisfies:
\[
\frac{d\pi_i}{df_e} = \frac{\partial f_e}{\partial \tau} \cdot \lambda(\tau) - \frac{dc_r}{df_e} = 0
\]

This yields a strategic pricing condition:
\[
f_e^* = \arg \max_{f_e} \left[ f_e \cdot \lambda(f_e) - c_r(f_e) \right]
\]
where \( \lambda(f_e) \) is the volume of transactions routed through edge \( e \) given fee \( f_e \), exhibiting demand elasticity.

\subsection{Characterisation of Outcomes}

\textbf{Monopoly Case:} In a topology where a single hub dominates all shortest paths, price-setting becomes unconstrained. Given inelastic demand for reachability, monopolistic hubs set:
\[
f_e^* = \max \left( \frac{\partial c_r}{\partial \tau} \bigg/ \frac{\partial \lambda}{\partial \tau},\; \text{reservation utility} \right)
\]
leading to rent extraction and enforced liquidity lock-in.

\textbf{Competitive Case:} When multiple equivalent paths exist, hubs undercut each other:
\[
f_e^* \to \min \left( c_r + \epsilon,\; \text{fee floor} \right)
\]
where \( \epsilon \) is the marginal strategic slack allowed by route churn.

\textbf{Liquidity Exhaustion Threshold:} At demand level \( D \) exceeding rebalancing capacity, liquidity fragmentation causes:
\[
\lim_{D \to D_c} \text{Routing Failure Rate} \to 1
\]
and hence:
\[
\lim_{D \to D_c} f_e^* \to \infty
\]

\subsection{Equilibrium Instability and Hysteresis}

When rebalancing costs exceed expected routing revenue, channels deplete and closure becomes rational:
\[
\mathbb{E}[f_e \cdot \lambda] < c_r \implies \text{Channel Abandonment}
\]

Under such conditions, the system exhibits hysteresis:
\[
\text{Post-failure recovery requires } \ell_e > \ell_e^{\text{crit}} \gg \ell_e^{\text{init}}
\]
indicating path-dependent failure and irreversibility in network structure.

This formalism underscores the structural brittleness of the Lightning routing economy under endogenous fee setting, bounded liquidity, and non-trivial transaction topology.

\newpage 

\section{Liquidity Topology Lemmas}

This section develops a formal characterisation of liquidity structures within the Lightning Network (LN) and their influence on routing capacity, economic stability, and agent incentives. We focus on deriving a set of lemmas that constrain the feasible topologies of channel liquidity in equilibrium and examine the consequences of liquidity consolidation, rebalancing thresholds, and hub-centric optimisation.

\subsection{Lemma 1: Liquidity Concentration Minimises Path Risk}

\textbf{Statement:} Let \( G = (V, E) \) be a directed channel graph with liquidity assignment \( \lambda: E \to \mathbb{R}_{\geq 0}^2 \). Then, for any path \( P \subset G \), the expected success rate of transaction routing increases monotonically with liquidity concentration in fewer nodes, i.e., hub centralisation.

\textbf{Proof Sketch:} Let \( P_1 \) and \( P_2 \) be two candidate paths of equal length and total liquidity, where \( P_1 \) includes multiple low-liquidity edges and \( P_2 \) routes through a single high-liquidity hub. Assuming independent failure probabilities \( p_i \sim \exp(-\ell_i) \) on each edge \( i \), the product of success probabilities satisfies:
\[
\prod_{i \in P_1} (1 - p_i) < \prod_{j \in P_2} (1 - p_j)
\]
because exponential decay penalises fragmentation. Hence, centralised liquidity reduces aggregate path risk.

\subsection{Lemma 2: Sparse Topologies Maximise Rent Capture}

\textbf{Statement:} In equilibrium, a sparsely connected LN topology dominated by a minimal set of high-volume hubs maximises fee extraction, subject to routing feasibility.

\textbf{Proof Sketch:} From the equilibrium condition \( f_e^* = \arg\max f_e \cdot \lambda(f_e) - c_r(f_e) \), monopoly pricing arises when alternative paths are infeasible or inefficient. By reducing edge redundancy and increasing betweenness centrality, hubs gain unilateral price-setting power. Thus, sparse topologies are rent-optimising under utility maximisation.

\subsection{Lemma 3: Full Rebalancing is NP-Hard in Arbitrary Topologies}

\textbf{Statement:} Let \( \Lambda(t) \) be the liquidity distribution across edges at time \( t \), and \( D(t) \) the demand matrix. The problem of finding an optimal rebalancing schedule to restore target liquidity \( \Lambda^* \) is NP-Hard.

\textbf{Proof Sketch:} The rebalancing task is reducible to the multi-commodity flow problem with path constraints, under time-evolving and asymmetric channel capacity. Known NP-hardness of multi-commodity flow with capacity and delay bounds (\cite{even1975complexity}) implies that global rebalancing in LN is computationally intractable in general networks.

\subsection{Lemma 4: Liquidity Starvation Threshold Exists}

\textbf{Statement:} For a given transaction rate \( \mu \) and average path length \( L \), there exists a critical liquidity level \( \ell_{\text{crit}} \) below which the probability of successful routing falls below \( \epsilon \), for any \( \epsilon > 0 \).

\textbf{Proof Sketch:} Let the routing success probability \( S(\ell) \sim 1 - e^{-k \cdot \ell} \) for some constant \( k > 0 \). Solving \( S(\ell) = \epsilon \) yields:
\[
\ell_{\text{crit}} = -\frac{1}{k} \ln(1 - \epsilon)
\]
demonstrating that when liquidity per edge falls below this threshold, routing reliability degrades exponentially, leading to service collapse.

These lemmas form the topological foundation for subsequent proofs regarding systemic failure, collapse of settlement dynamics, and entrenchment of hub dominance.

\newpage 
\section{Glossary of Symbols}

This section provides a comprehensive reference of the mathematical symbols and structures used throughout the paper, ensuring clarity and unambiguous interpretation in all formal derivations and game-theoretic constructions.

\begin{longtable}{p{3cm} p{12cm}}
\toprule
\textbf{Symbol} & \textbf{Definition} \\
\midrule
\( \mathscr{S} \) & The full system tuple: \( \mathscr{S} = (\mathscr{B}, \mathscr{L}, \mathscr{A}, \mathscr{R}, \mathscr{V}, \mathscr{C}, \mathscr{T}) \) \\
\( \mathscr{B} \) & Base-layer blockchain system (BTC ledger with limited throughput) \\
\( \mathscr{L} \) & Overlay network, instantiated as the Lightning Network \\
\( \mathscr{A} \) & Set of agents: users \( \mathcal{U} \), hubs \( \mathcal{H} \), watchtowers \( \mathcal{W} \), miners \( \mathcal{M} \) \\
\( \mathscr{R} \) & Routing functions: mapping from transaction intents to viable payment paths \\
\( \mathscr{V} \) & Transaction valuation space (subjective utility values per transaction) \\
\( \mathscr{C} \) & Cost function space, encompassing on-chain and off-chain penalty structures \\
\( \mathscr{T} \) & Discrete temporal index space (epochs over which system evolves) \\
\( \mathcal{B} \) & Set of all blockchain blocks \\
\( \sqsubseteq \) & Total order over blocks induced by Proof-of-Work consensus \\
\( \delta \) & Inter-block interval function: \( \delta: \mathscr{T} \to \mathbb{N} \) \\
\( s \) & Block size limit (bytes or virtual bytes) \\
\( \bar{t}_{\text{tx}} \) & Mean transaction size (bytes) \\
\( T_{\max} \) & Maximum theoretical base-layer transaction throughput \\
\( V(t) \) & Active nodes in Lightning Network at time \( t \) \\
\( E(t) \) & Active channels in Lightning Network at time \( t \) \\
\( \lambda(e_{ij}) \) & Liquidity available for edge \( e_{ij} \), defined as \( (\ell_{ij}, \ell_{ji}) \) \\
\( \sigma_k \) & Strategy function for agent \( a_k \), maximising utility over time \\
\( u_k(t) \) & Utility of agent \( a_k \) at epoch \( t \) \\
\( \delta \in (0,1] \) & Discount factor for utility valuation over time \\
\( \mathcal{P} \) & Set of all simple routing paths in \( \mathscr{L} \) \\
\( \rho(P) \) & Path validity function: \( \rho(P) = 1 \) if constraints are satisfied \\
\( \phi(P) \) & Cost function associated with path \( P \) \\
\( P^* \) & Optimal routing path under constraint-aware minimisation \\
\( \tau \) & Transaction (element of \( \Pi(t) \)) \\
\( v(\tau) \) & Utility value of transaction \( \tau \) \\
\( \Pi(t) \) & Set of transactions initiated at time \( t \) \\
\( U_{\text{net}}(t) \) & Aggregate transactional utility at time \( t \) \\
\( \mathscr{C}_{\mathscr{B}} \) & Cost subspace: mempool congestion, miner fees, delays \\
\( \mathscr{C}_{\mathscr{L}} \) & Cost subspace: routing penalties, liquidity inefficiencies, rebalancing \\
\( \ell \) & Liquidity quantity associated with a channel or path \\
\( \mu \) & Mean transaction rate (TPS) across the network \\
\( L \) & Mean effective routing path length (hops) \\
\( \Lambda(t) \) & Liquidity distribution matrix at time \( t \) \\
\( D(t) \) & Transaction demand matrix at time \( t \) \\
\( \ell_{\text{crit}} \) & Minimum required liquidity per edge to sustain system function \\
\( S(\ell) \) & Success probability of routing given liquidity level \( \ell \) \\
\( G = (V, E) \) & Directed Lightning graph with vertex set \( V \) and edge set \( E \) \\
\bottomrule
\end{longtable}

\end{document}